\documentclass[preprint,11pt]{elsarticle}
\usepackage[english]{babel}
\usepackage{booktabs}
\usepackage{adjustbox}
\usepackage{blindtext}
\usepackage{hyperref}
\usepackage{rotfloat}
\usepackage{hyperref}
\usepackage{natbib}
\usepackage{bbm}
\usepackage{bm}
\usepackage{amsmath}
\usepackage{amssymb}
\usepackage{subcaption}
\usepackage{caption}
\usepackage{float}
\begin{document}
\begin{frontmatter}
\title{\textbf{Comparative Analysis of Richardson-Lucy Deconvolution and Data Unfolding with Mean Integrated Square Error Optimization}}
\author{Nikolay D.\ Gagunashvili\corref{author}}
\ead{nikolay@hi.is}
\address{University of Iceland, S\ae mundargata 2, 101 Reykjavik, Iceland}
\begin{abstract}
Two maximum likelihood-based algorithms for unfolding or deconvolution are considered: the Richardson-Lucy method and the Data Unfolding method with Mean Integrated Square Error (MISE) optimization. Unfolding is viewed as a procedure for estimating an unknown probability density function. Both external and internal quality assessment methods can be applied for this purpose. In some cases, external criteria exist to evaluate deconvolution quality. A typical example is the deconvolution of a blurred image, where the sharpness of the restored image serves as an indicator of quality. However, defining such external criteria can be challenging, particularly when a measurement has not been performed previously. In such instances, internal criteria are necessary to assess the quality of the result independently of external information.
The article discusses two internal criteria: MISE for the unfolded distribution and the condition number of the correlation matrix of the unfolded distribution.
These internal quality criteria are applied to a comparative analysis of the two methods using identical numerical data. The results of the analysis demonstrate the superiority of the Data Unfolding method with MISE optimization over the Richardson-Lucy method.

\end{abstract}
\begin{keyword}
probability density function estimation\sep
cluster analysis \sep 
inverse problem \sep 
system identification  \sep
entropy regularization  
\end{keyword}
\end{frontmatter}
\vspace{2cm}
\section {Introduction}
Unfolding, or deconvolution, is a common part of data analysis used to improve the resolution of measurements and remove distortions related to inefficiencies in event registration. Computer modeling is often employed to study the registration process.\\

As discussed  in \cite{ng}, the data under analysis can be  represented through two sets of random variables:
\begin{enumerate}
\item The measured data sample is a collection of  random variables:
\begin{equation}
 y_1, y_2,\ldots ,y_{n} \label{meas}
\end{equation}
with the  Probability Density Function (PDF) $f (y)$.
\begin{itemize}
\item The true PDF  $\phi(x)$ is not known.
\item The measured PDF $f(y)$  is also unknown, however, it can be estimated.
\end{itemize}
\item The simulated data sample is a collection of  pairs of random variables:
\begin{equation}
x_1^s, y_1^s ; \, x_2^s, y_2^s; \, \ldots \,x_k^s, y_k^s  \label{simul}   
\end{equation}
with  PDF's  $\phi^s(x)$ and  $f^s(y)$. 
\begin{itemize}
\item The generated PDF  $\phi^s(x)$ is an analog of the true PDF $\phi(x)$.  
In the common case,  $\phi^s(x) \ne \phi (x)$ and the PDF $\phi^s(x)$  can be estimated. 
\item The reconstructed PDF $f^s (x)$  is an analog of the measured PDF $f(x)$ and  the PDF   $f^s (x)$ can be estimated. 
 \end{itemize}
\end{enumerate}
The simulated sample serves the purpose of constructing a mathematical model for the measurement system. 

Unfolding in High Energy Physics, Nuclear Physics, Particle Astrophysics and Radiation Protection Physics   often employs the histogram-based  method to estimate PDF.
 The most popular and easily interpretable unfolding methods, as discussed in \cite{review}, include the Richardson-Lucy method \cite{taras, Richardson, Lucy}, the Singular Value Decomposition method \cite{kart}, the Tikhonov regularization method  \cite{Tikhonov77}, and the  Bin-by-Bin correction factor method  \cite{bintobin}. 
These methodologies generally operate under the assumption that the measurement process does not introduce non-linear distortions to the observed distribution.

 A commonly used model for this scenario is the Fredholm integral equation, which described  the relationship between the measured PDF denoted as $f(y)$ and the true PDF  denoted as $ \phi(x)$.

\begin{equation}
\int \limits_{-\infty}^{+\infty} R(x,y)A(x) \phi(x) dx=f(y),  \,\,    \label{fred}
\end {equation}
where $A(x)$ is the probability of recording of an event with a characteristic $x$
(the acceptance); $R(x,y)$, is the probability of obtaining $y$ instead of $x$ (the
experimental resolution).
The linear  dependence of the measured PDF $f(y)$ on the true PDF $\phi(x)$  is reasonable, nevertheless in many cases 
application of the Fredholm integral  equation (\ref{fred})  is not obvious.\\ Let us define, in the histogram approach,  the binning $b_1,b_2,...b_{l+1}$ for the measured PDF  $f(y)$ and the binning $a_1, a_2,....a_{m+1}$  for the true PDF  $\phi(x)$.
Dependence of the measured distribution $f(y)$ from the true distribution $\phi(x)$  is approximated by a system of linear equations:
\begin{equation}
(f_1, f_2,...., f_l)^T= \mathbf {R } \times ( \phi_1,  \phi_2,..., \phi_m)^T,
\end{equation}
where  $f_i= \int\limits_{b_i}^{b_{i+1}} f(y)dy$,  $\phi_i= \int\limits_{a_i}^{a_{i+1}} \phi(x)dx$  and $\mathbf {R }$ is the response matrix, or
the linear approximation of the mathematical model of the measurement system. 
\section {Quality assessment  of unfolding  method}
 Both external and internal criteria are employed to assess the quality of the unfolding procedure. In certain instances, external criteria are available, providing a framework for evaluating the reliability of the procedure. However, in experimental physics, defining external criteria can be challenging, particularly when no prior measurements exist. In such scenarios, model-based assumptions often serve as a guide. Conversely, internal criteria for evaluating the quality of the unfolding result become essential when external references are unavailable.

The quality criteria for the data unfolding procedure, as presented below, are integral to the evaluation process. Notably, these criteria not only aid in determining the optimal value of the regularization strength in unfolding algorithms but also facilitate the comparison of different algorithms.

\vspace{2cm}
\subsection{Accuracy}
One commonly used measure of accuracy in the estimation of a PDF  $\hat{\phi}(x)$  is the Mean Integrated Square Error (MISE) \cite{silver, fridman,ng}, defined as:

\begin{equation}
\mathrm{MISE} =\int \limits_{-\infty}^{+\infty} \mathbf {E}[(\hat{\phi}(x)-\phi (x))^2]dx
\end{equation}
For the histogram estimation of the unfolded distribution  
\begin{equation}
\hat{\phi}(x)=\frac {\hat{\phi_i}}{a_{i+1}-a_{i}} \, \,\text{for} \, \, \, a_{i} \leq x <  a_{i+1}, 
\end{equation}
were $\hat{\phi_i}$,  is  an estimator  of  $\phi_i$.
The  MISE in this case is expressed as:
\begin{equation}
 \begin{aligned}
\mathrm{MISE} = & \int \limits_{a_1}^{a_{m+1}}\mathbf{E} \,[\hat{\phi }(x)^2] dx- 2 \int \limits_{a_1}^{a_{m+1}} \phi(x)  \mathbf{E}[\hat{\phi }(x)] dx+  \int \limits_{a_1}^{a_{m+1}}\phi (x)^2dx\\
=&\mathbf{E} \sum_{i=1}^m \frac {\hat{\phi_i}^2} { a_{i+1}-a_{i}} 
- 2  \mathbf{E} \sum_{i=1}^m   \frac {\hat{\phi_i}} { a_{i+1}-a_{i}}    \int \limits_{a_i}^{a_{i+1}} \phi (x)dx + \int \limits_{a_1}^{a_{m+1}}\phi (x)^2dx. \label{msem}
\end{aligned}
\end{equation}
The last term of the expression  (\ref{msem}) does not depend on the estimators  $\hat{p_i}$   (it is a constant). Therefore, the  choice of an optimal parameter with the minimal value of $ \mathrm{MISE}$ does not depend on this part of equation.
\subsection{Condition number}
Another measure of the quality of the unfolding process is the condition number of the correlation matrix for $\hat{\phi_i}/( a_{i+1}-a_{i}) $.  
The estimators of probabilities $\hat{\phi_i}$  satisfy the  equation  $\sum_{i=1}^m \hat{\phi_i}=1$ and as a result, the  correlation matrix is often nearly  singular. 

To address this, it is advisable to identify the minimum condition number (MCN) of the correlation matrix when excluding one bin from consideration.
\begin{equation}
\mathrm{MCN} =  \underset{k}{\text{argmin}}[ \mathrm{COND} (C_{-k})] ,
\end{equation}
where $C_{-k}$ is full correlation matrix without bin $k$.

\section{Unfolding methods used for comparison}
Two multidimensional methods for the optimization of the likelihood function are compared:
\begin{itemize}
\item{ Richardson-Lucy  method (RL)  is an iteration method, where the number of iterations is used as a regularization parameter. }
\item{New data unfolding with Mean Integrated Squared Error optimization (NG) proposed by the author of this paper \cite{ng}.
It is a method   with entropy  regularization.}
\end {itemize} 
For both investigated unfolding methods, binning was defined:
\begin{itemize}
\item{Equidistant bins are used for the RL method  for the true and reconstructed histograms.}
\item {K-mean clustering is used  for  NG method   for the true and reconstructed histograms.}
\end {itemize} 
The response Matrix $R$ is calculated with simulated data:
\begin{itemize}
\item{ The standard  method  for the calculation of matrix elements in the RL method is used according to Equation 3 from  \cite{review}.} 
\item {The system identification method described in  \cite{ng} is used for the calculation of matrix elements  in the NG method.}
\end{itemize}
For each  case, we can calculate the unfolded distribution, the correlation matrix and the average unfolded distribution for the minimal value of  MISE and  the average unfolded distribution with the minimum MCN. 

\section{Numeric example}
Two types of data were  simulated:  data for the application unfolding algorithm and data for the identification system or calculation of the $R$ matrix.
\subsection{ Data for unfolding}
\begin{figure}
\vspace{-2cm} 
	\centering
	\begin{subfigure}{0.495\linewidth}
		\includegraphics[width=\linewidth]{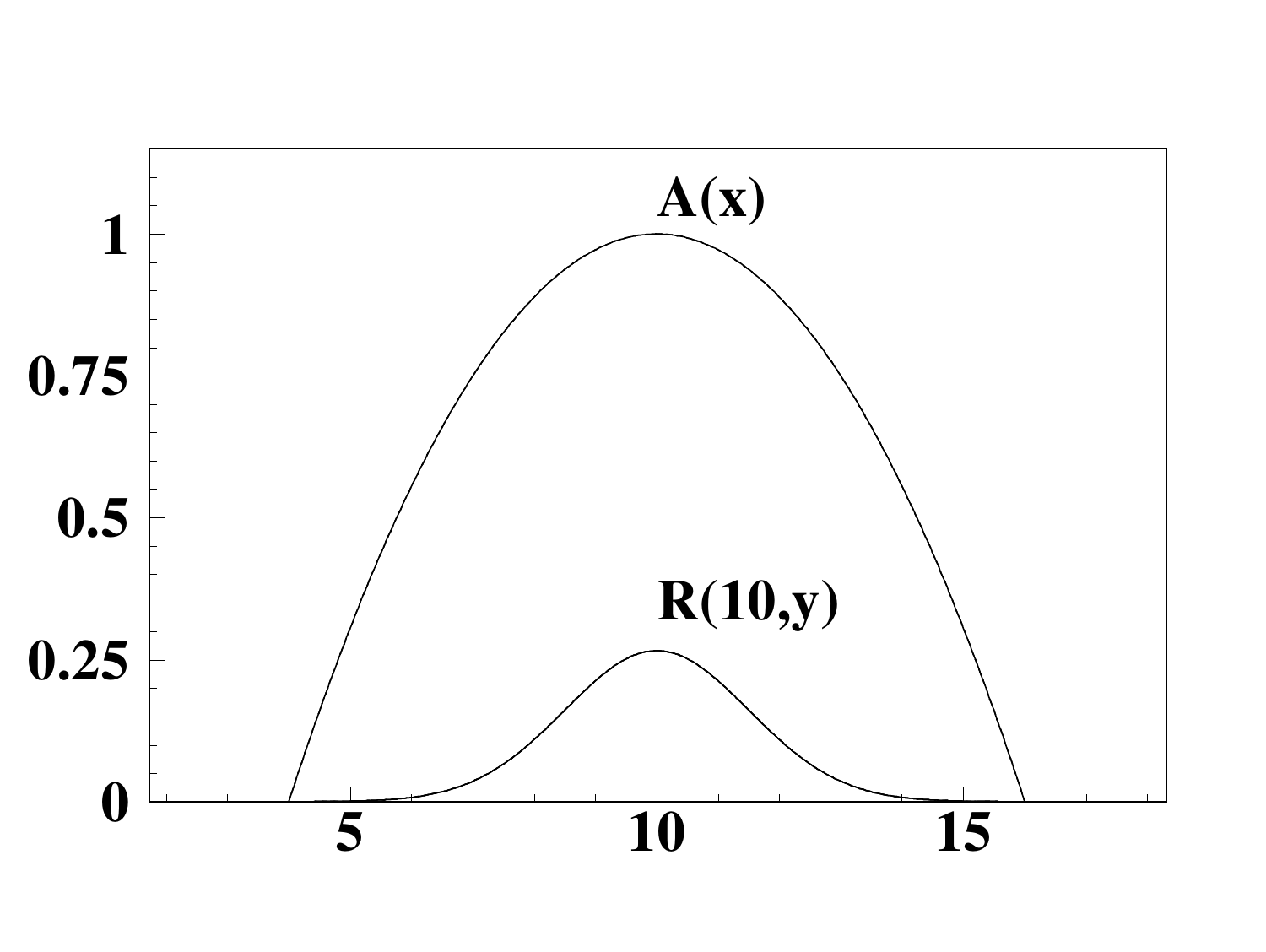}
	\end{subfigure}
	\caption{Acceptance $A(x)$\/ and resolution function $R(x,y)$ ,   for $x=10$ .} 
	\label{fig:accept}
\end{figure}

Following to \cite{zhig} let us assume a true PDF $\phi(x)$  that is described by a sum of two Breit-Wigner functions defined on the interval $[4, 16]$:
\begin{equation}
 \phi(x) \propto 2\frac{1}{(x-10)^2+1} + \frac{1}{(x-14)^2+1}  ,
\label{testform}
\end{equation}
from which the measured PDF $f(y)$  is obtained according to equation (\ref{fred})
with the acceptance function $A(x)$ (Figure~\ref{fig:accept}):
\begin{equation}
A(x)=1-\frac{(x-10)^2}{36}  , \label{acc}
\end{equation}
and the resolution function describing Gaussian smearing (Figure~\ref{fig:accept}):
\begin{equation}
R(x.y)=\frac{1}{\sqrt{2\pi}\sigma}\exp\left(-\frac{(y-x)^2}{2\sigma^2}\right), \, \sigma=1.5\; \label{res}  .
\end{equation}
The measured distribution obtained by
simulating   events according to the PDF $\phi(x)$  is  shown in Fig.~\ref{fig:true}. 
Two examples  were considered with 10000 events  and 1000 events simulated according formula (\ref{testform})
\subsection{Data for system identification}
\begin{figure}
\vspace*{-2cm}
	\centering
 	\begin{subfigure}{0.495\linewidth}
		\includegraphics[width=\linewidth]{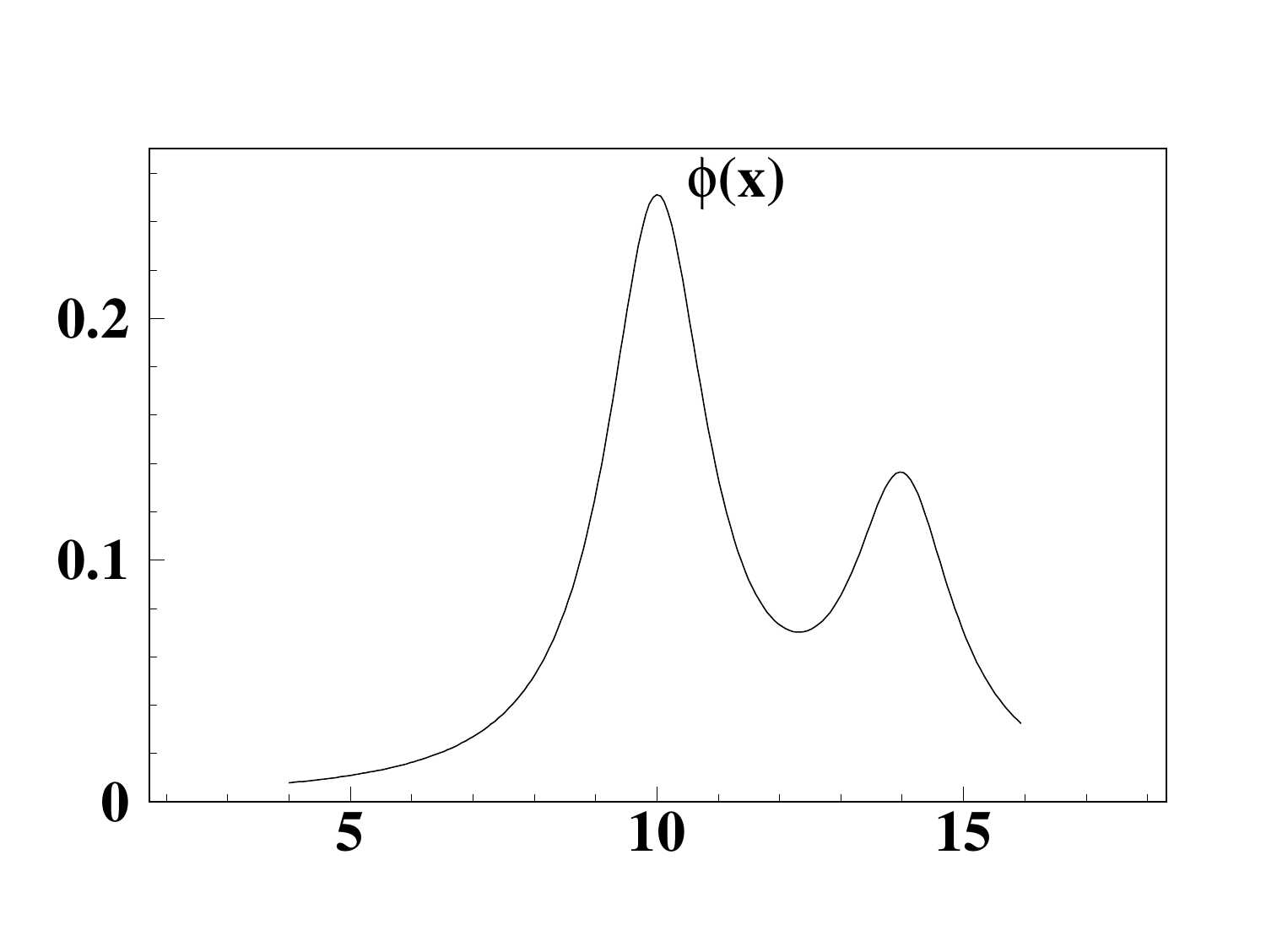}
	\end{subfigure}
   \hfill 
	\begin{subfigure}{0.495\linewidth}
		\includegraphics[width=\linewidth]{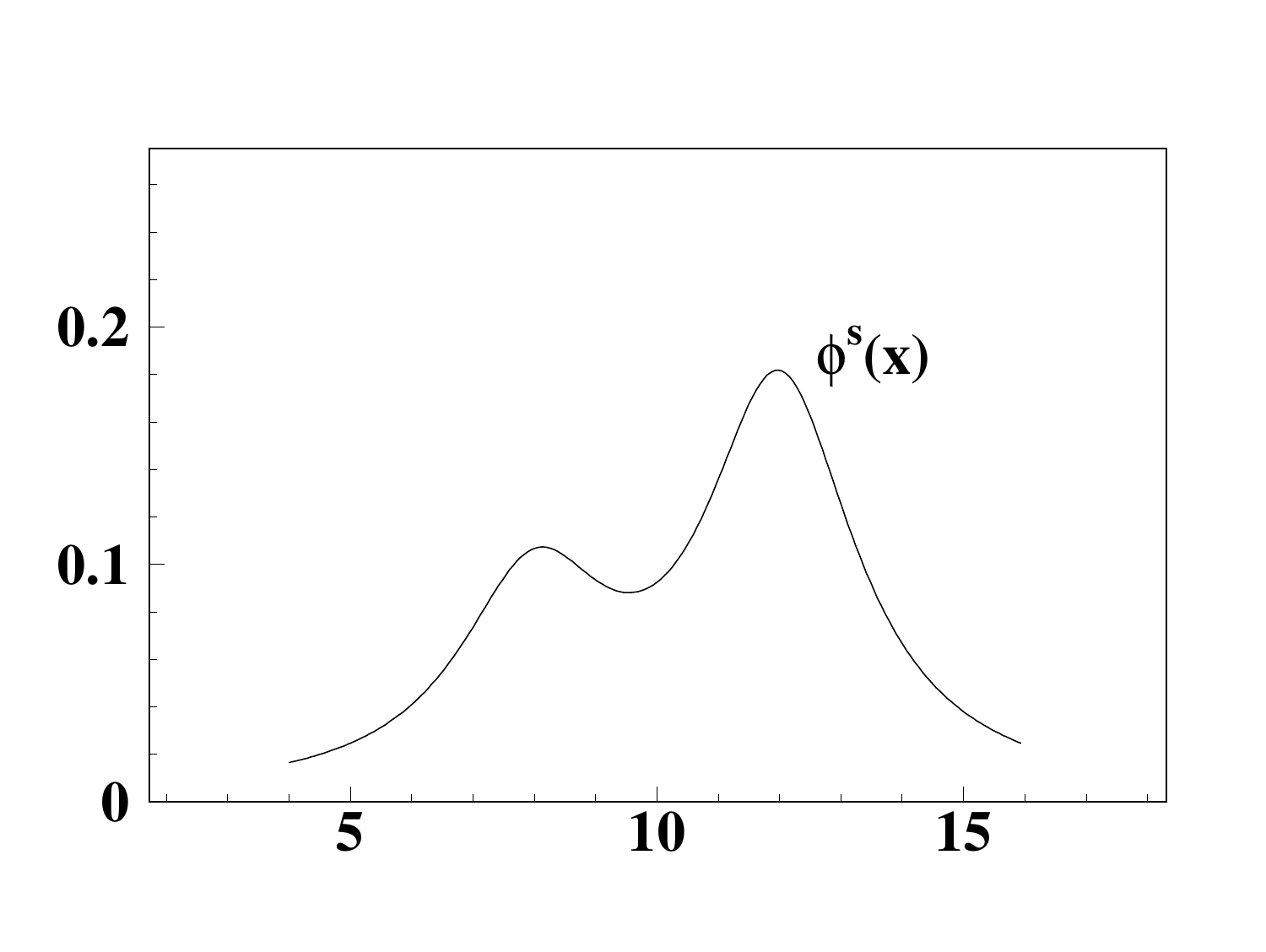}
	\end{subfigure}        
	\caption{True  PDF $\phi (x)$ (left) , PDF used for simulation   $\phi^s(x)$  (right).}
	\label{fig:true}
\end{figure}

To create a model of  the measurement system, we simulate  it as described  by a sum of two Breit-Wigner functions defined as:
\begin{equation}
 \phi^s (x) \propto \frac{1.5^2}{(x-8)^2+1.5^2} + 2\frac{1.5^2}{(x-12)^2+1.5^2}  \label{rcs}  .
\end{equation}
From this, the reconstructed  PDF $f^s(y)$  is obtained according to equation (\ref{fred}) with the same acceptance  $A(x)$  equation  (\ref{acc}) and resolution function  $R(x,y)$  (\ref{res}).
The reconstructed distribution, obtained by simulating   $10^6$  events according to PDF $\phi^s(x)$,  is  shown in Figure~\ref{fig:true}.
Notice that the distribution used for the creation of the model of the measurement system in Figure~\ref{fig:true} (right)  essentially differs from the distribution used for the simulation of the measured data in  Figure~\ref{fig:true} (left)
\newpage
\subsection{Unfolding results for the case of 10000 events.}
Twenty equidistant bins are defined for the unfolded distribution and 40 equidistant bins for the histogram of reconstructed events in the RL case. Using K-means clustering, 20 bins are defined for the unfolded distribution and 40 bins for the histogram of reconstructed events in the case of the NG method.

\begin{figure}[h]
	\centering
 	\begin{subfigure}{0.495\linewidth}
		\includegraphics[width=\linewidth]{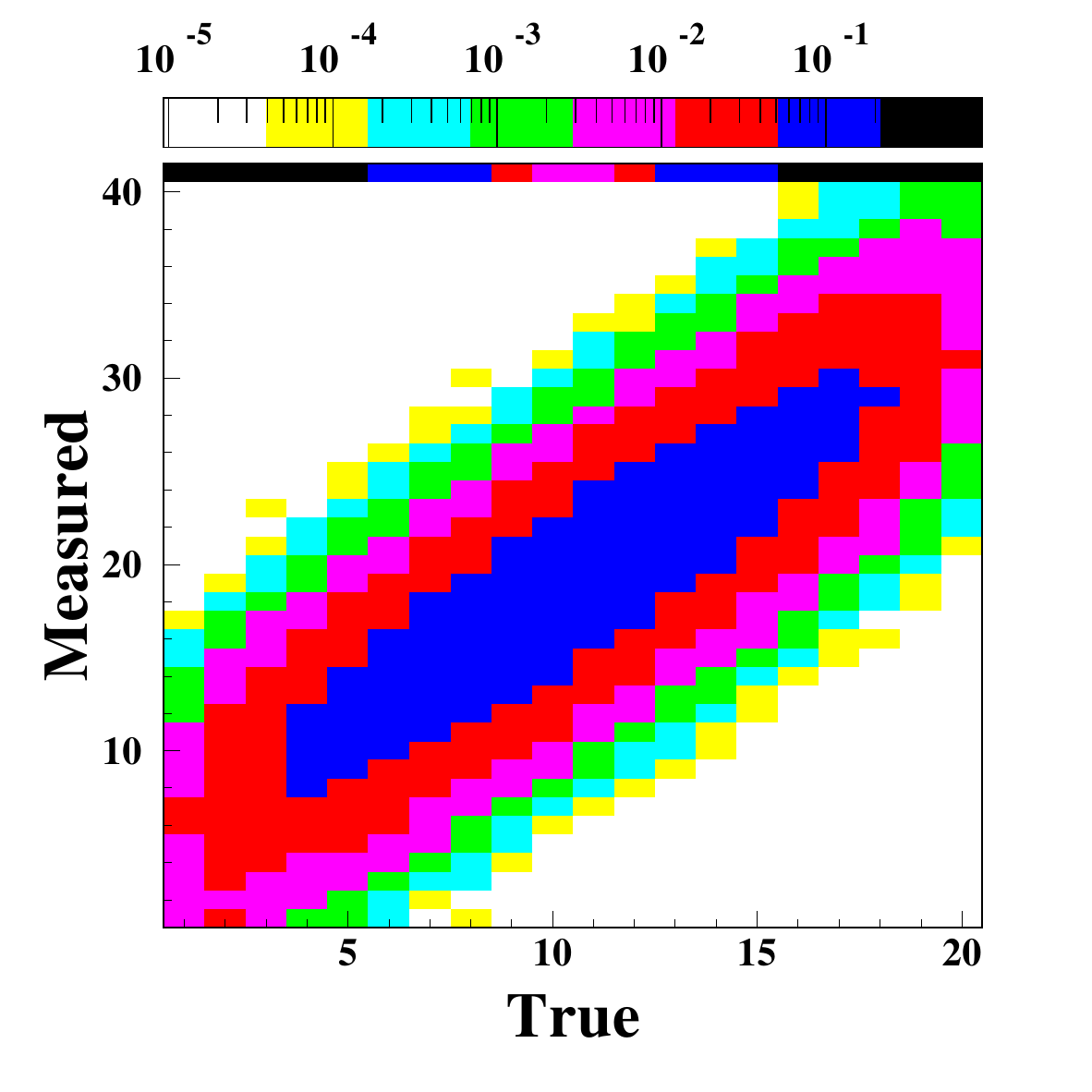}
	\end{subfigure}
   \hfill 
	\begin{subfigure}{0.495\linewidth}
		\includegraphics[width=\linewidth]{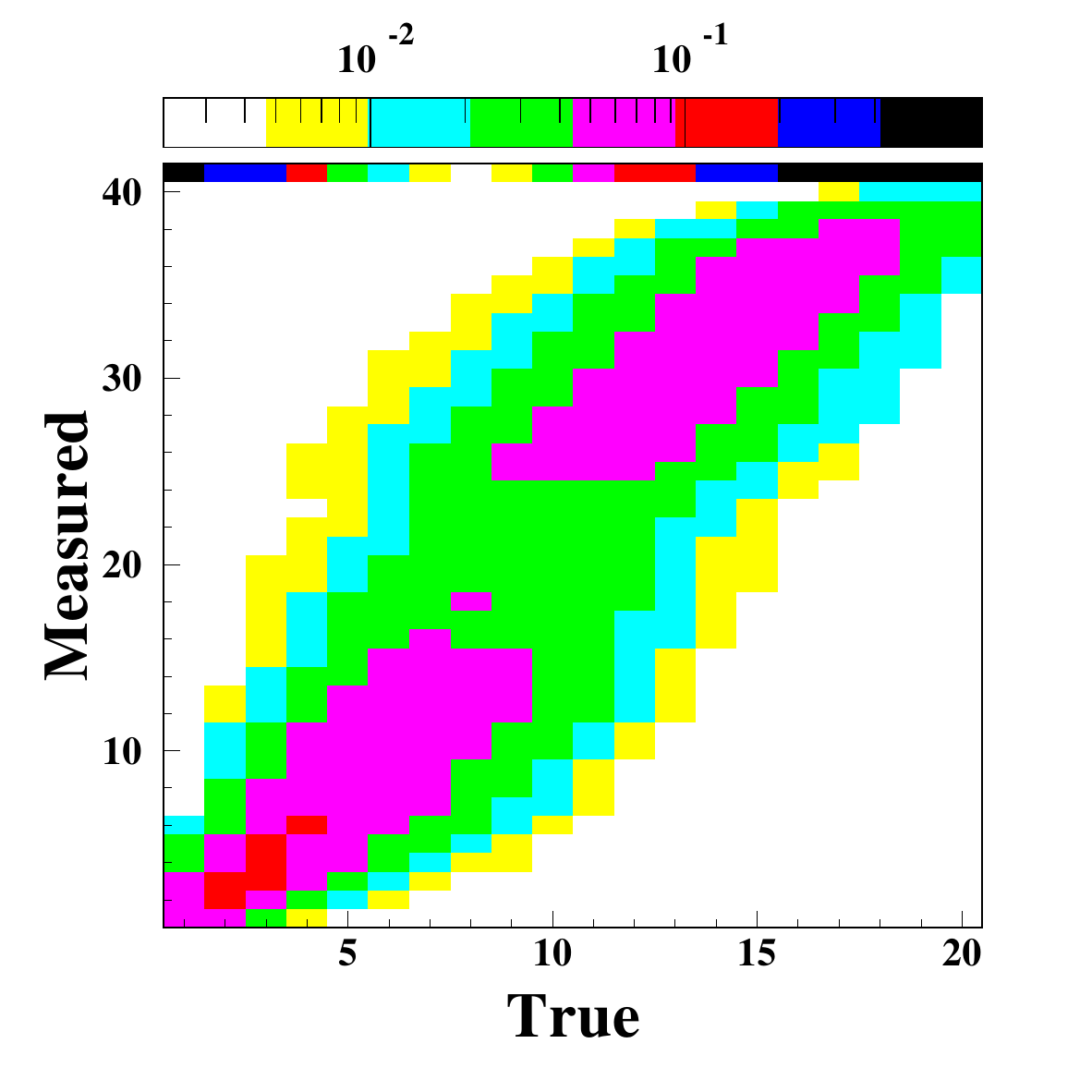}
	\end{subfigure}   
         \caption{ Matrix $R$ calculated for RL method (left) and NG method (right).}     
	\label{fig:rmatr}
\end{figure}

\begin{figure}
	\centering
 	\begin{subfigure}{0.495\linewidth}
		\includegraphics[width=\linewidth]{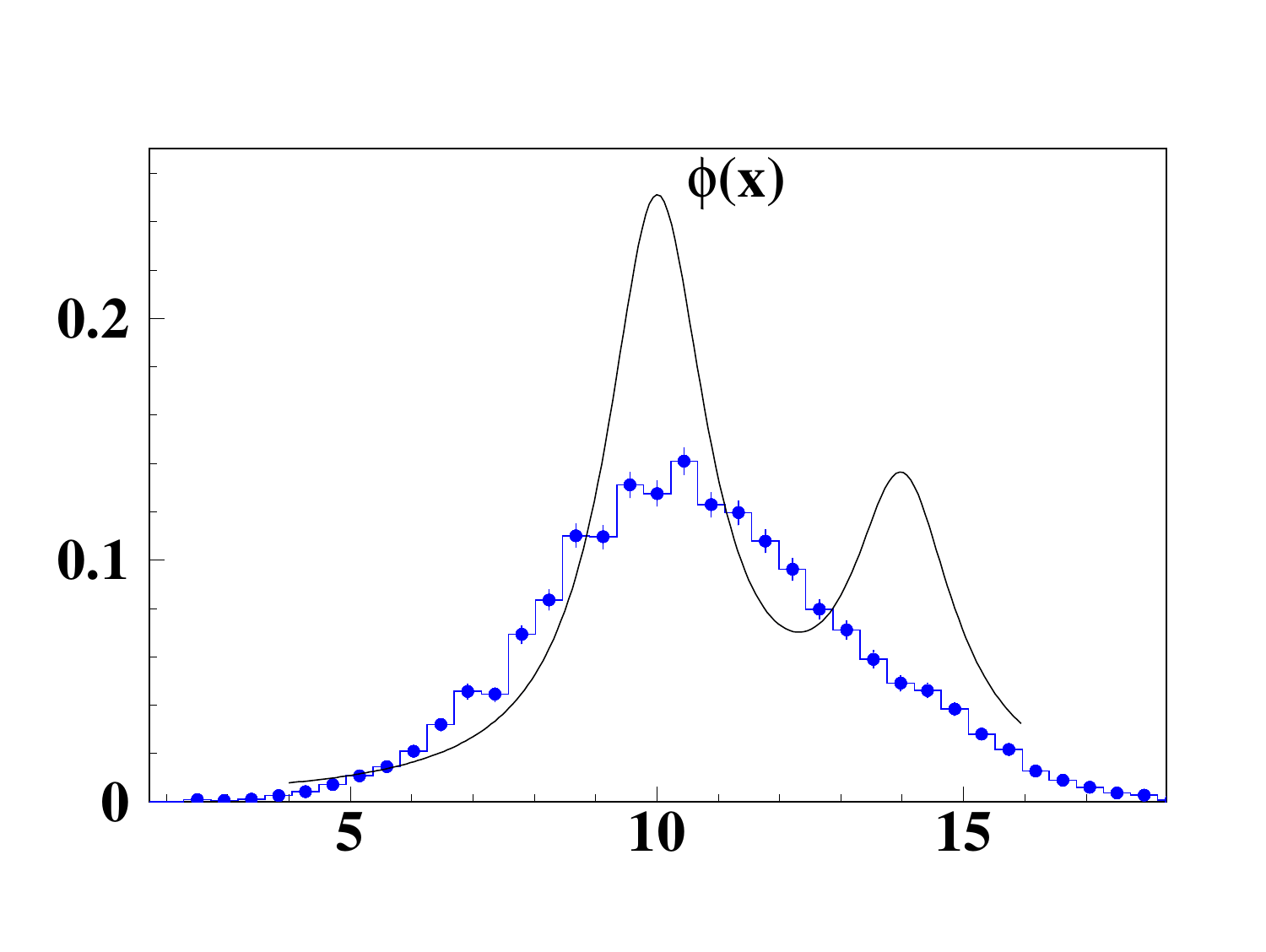}
	\end{subfigure}
   \hfill 
	\begin{subfigure}{0.495\linewidth}
		\includegraphics[width=\linewidth]{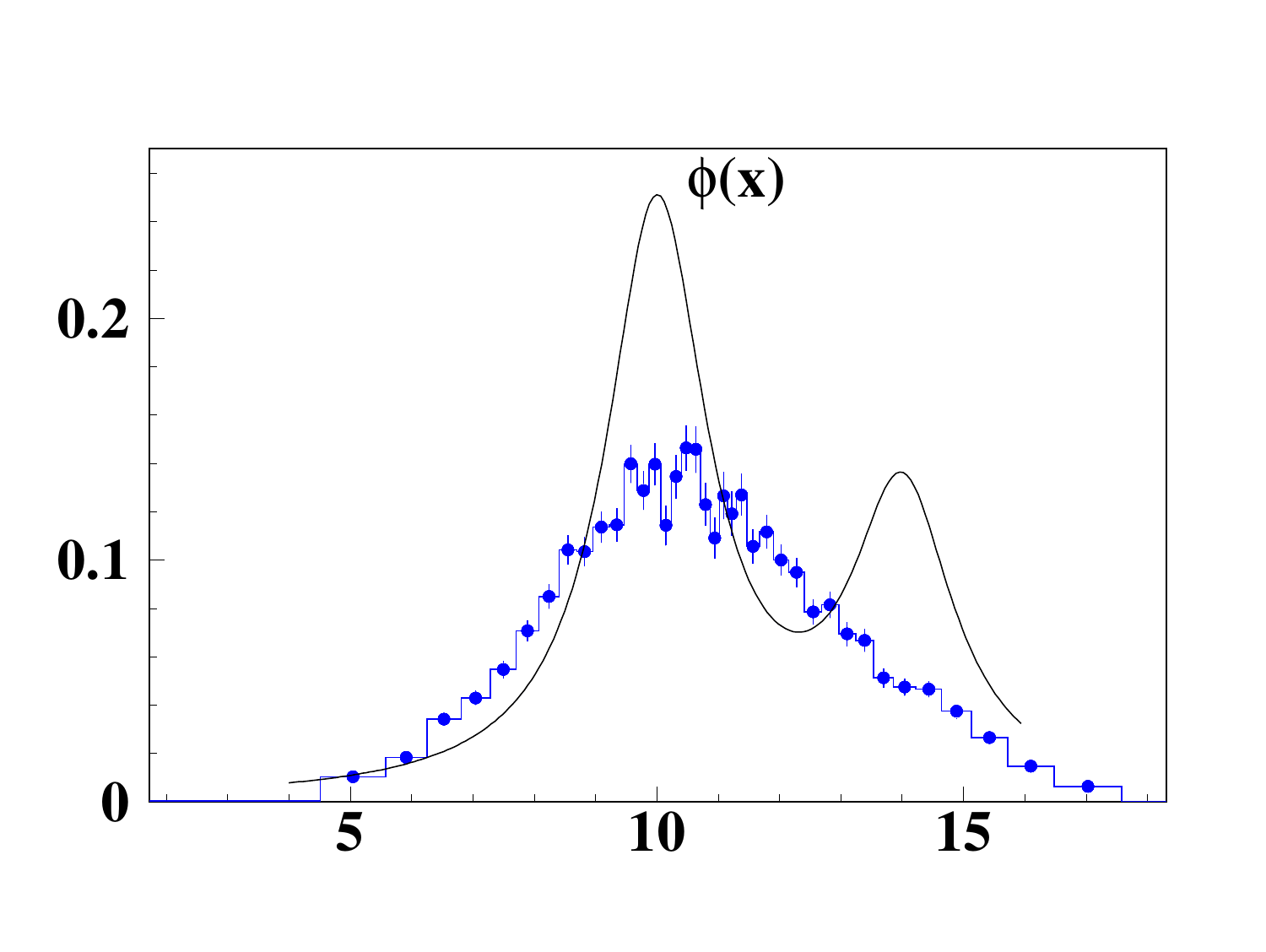}
	\end{subfigure}        
	\caption{Measured  distribution histogram for the RL method (left)  and measured distribution histogram for NG method (right). The true distribution  is shown  by the curve.}
	\label{fig:meas}
\end{figure}

\begin{figure}
	\centering
 	\begin{subfigure}{0.495\linewidth}
		\includegraphics[width=\linewidth]{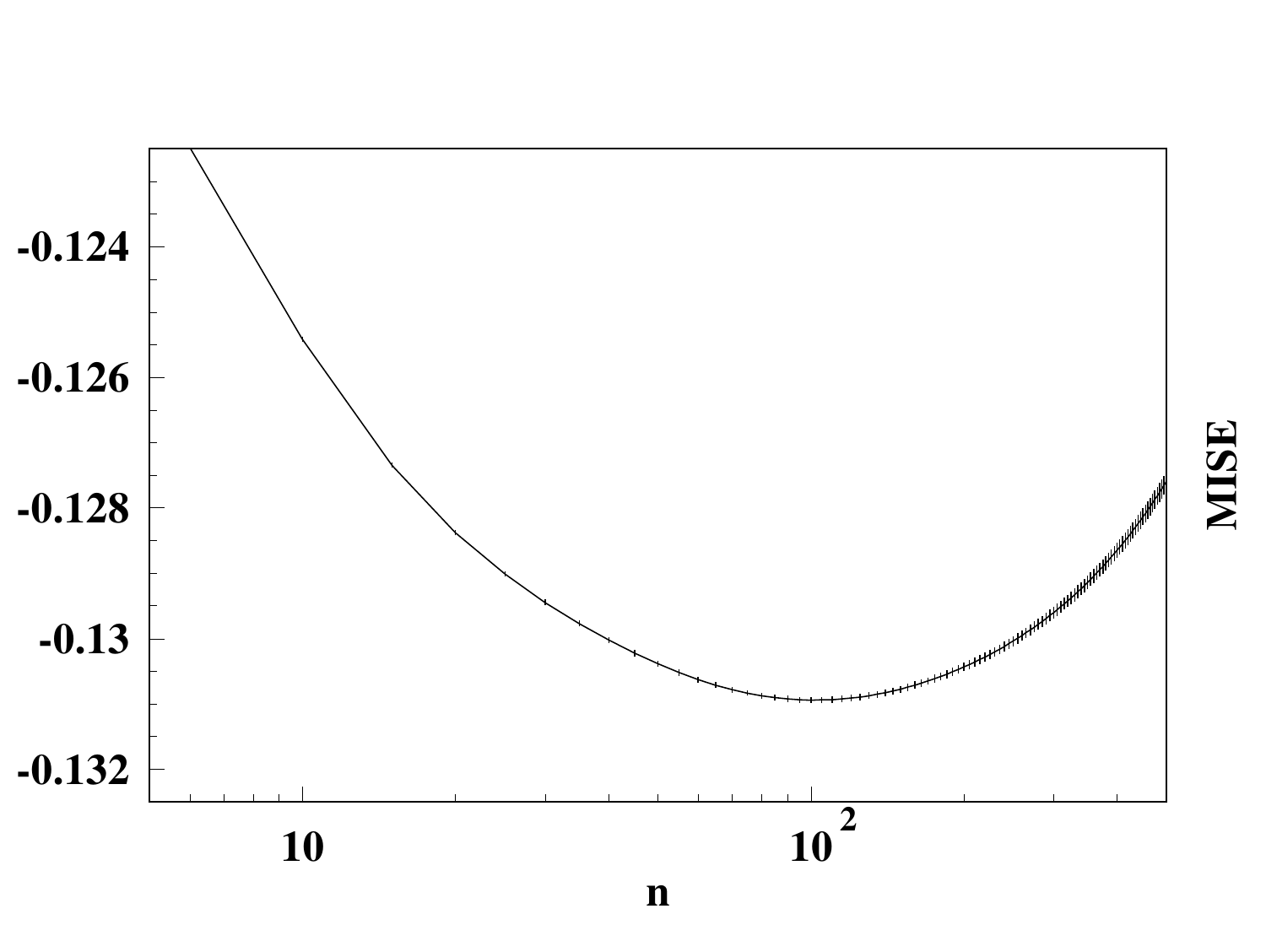}
	\end{subfigure}
   \hfill 
	\begin{subfigure}{0.495\linewidth}
		\includegraphics[width=\linewidth]{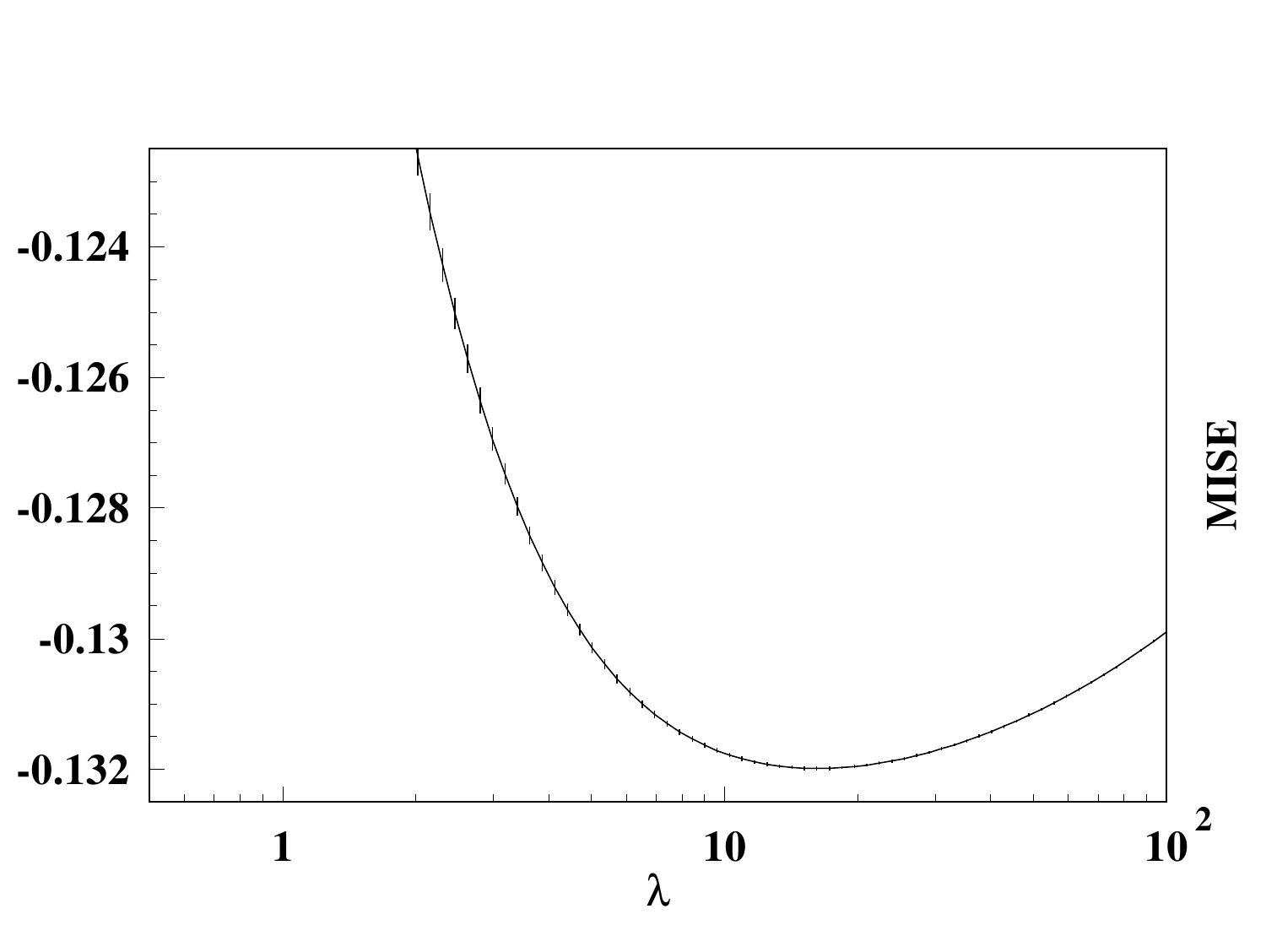}
	\end{subfigure}        
	\caption{MISE for RL method (left) and MISE for the NG method(right).  The standard deviations of the estimated MISE are presented as error bars.
  }
	\label{fig:mise1}
\end{figure}

\begin{figure}
	\centering
 	\begin{subfigure}{0.495\linewidth}
		\includegraphics[width=\linewidth]{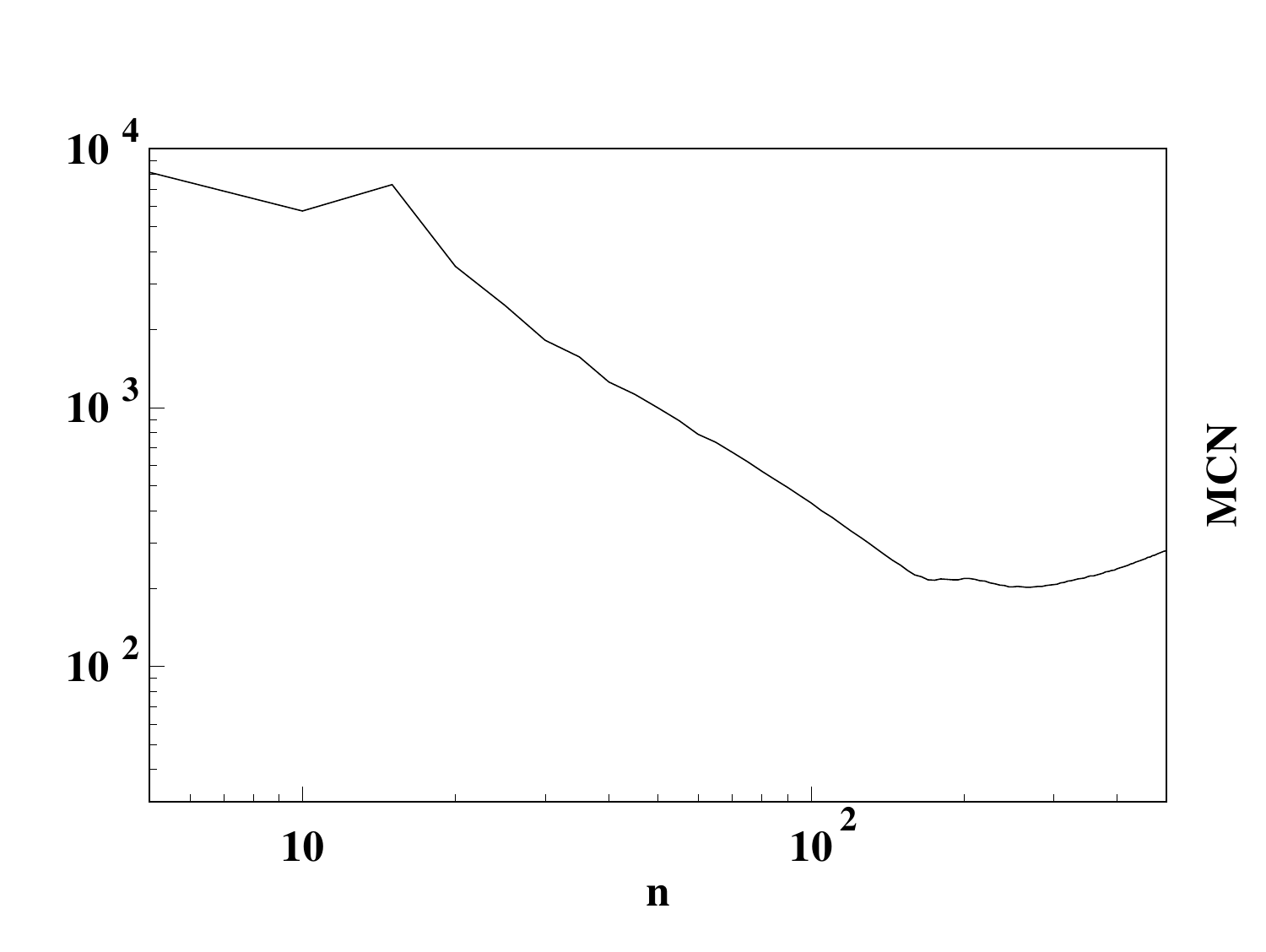}
	\end{subfigure}
   \hfill 
	\begin{subfigure}{0.495\linewidth}
		\includegraphics[width=\linewidth]{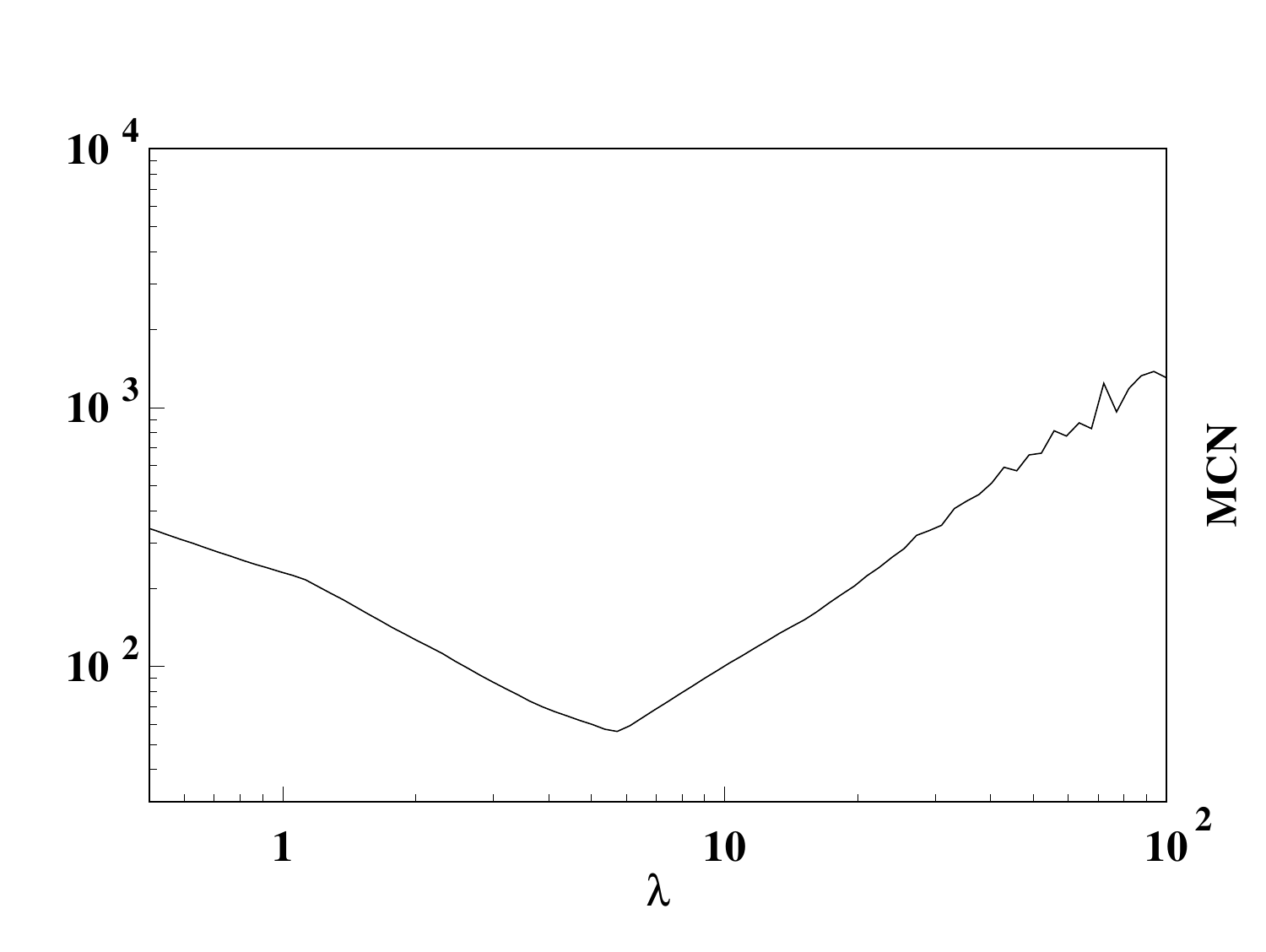}
	\end{subfigure}        
	\caption{MCN for RL method (left) and MCN for NG method (right). }
	\label{fig:cond1}
\end{figure}

The response matrices are presented in Figure~\ref{fig:rmatr}, and the histograms of measured distributions are shown in Figure~\ref{fig:meas}. To determine the minimum values of the MISE and of the MCN, the regularization parameter was scanned in both cases. The results of the calculations are presented in Figure~\ref{fig:mise1} and  \ref{fig:cond1}. Five hundred replications of the unfolded distribution were used for this analysis.

\newpage
\vspace*{-3 cm}
\begin{center}
\captionof{table}{Characteristics of the minima of MISE and  MCN for the  RL  and NG algorithms for case of 10000 measured events.}
\begin{tabular}{ | c | l l  l | l  l  l |  }
\hline
& \multicolumn{3}{ c }{Minimum MISE }  & \multicolumn{3} {c|} {Minimum MCN} \\
\hline
 Method &MISE & MCN & Reg. par. &MISE & MCN & Reg. par. \\
\hline 
  RL & -0.13094&416.6 &102  & -0.12994 &203.1 &262\\
  NG & -0.13199 &162.1  & 16.19 &-0.13062  & 56.0 & 5.72\\
\hline
\end{tabular}
\end{center}

According  to the data presented in Table 1, the NG method exhibits a lower value of the optimal MISE in both cases. It also has a lower MCN compared to the RL method, indicating that the NG method handles multicollinearity more effectively. This suggests that the NG method is likely to be more robust and produce more stable results.

\begin{figure}[h]
	\centering
 	\begin{subfigure}{0.495\linewidth}
		\includegraphics[width=\linewidth]{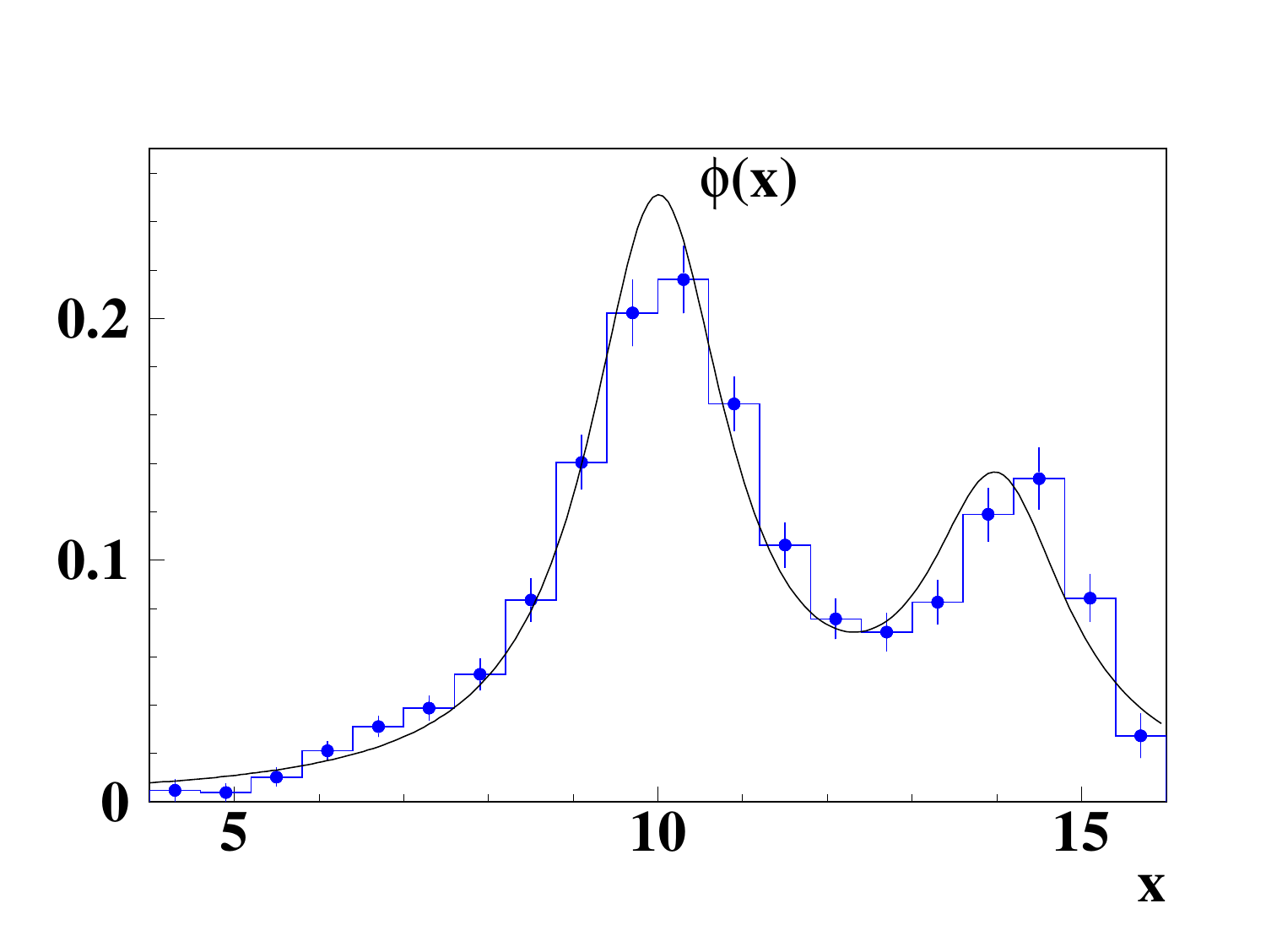}
	\end{subfigure}
   \hfill 
	\begin{subfigure}{0.495\linewidth}
		\includegraphics[width=\linewidth]{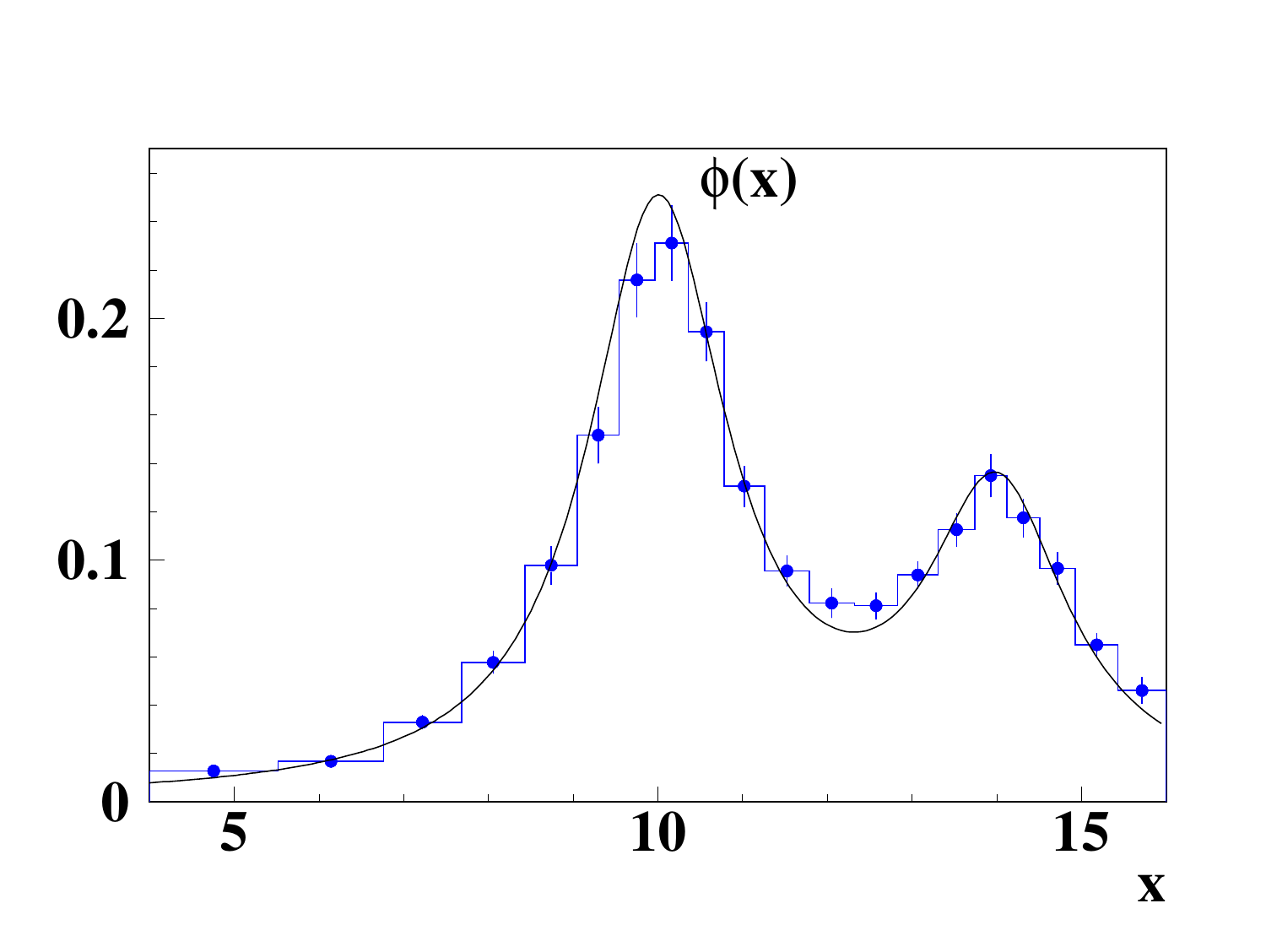}
	\end{subfigure}        
	\caption{Unfolded PDF  of the RL method with the minimum MISE value (left)  and the unfolded PDF of the NG method with  the minimum MISE value (right). The true distribution  is represented  by the curve. For bin $i$, the error bar represents the standard deviation  of  $\hat{\phi_i}/(a_{i+1}-a_{i})$.}
\label{fig:res2}
\end{figure}
\begin{figure}[h]
     \vspace *{-2cm}
	\centering
 	\begin{subfigure}{0.495\linewidth}
                \includegraphics[width=\linewidth]{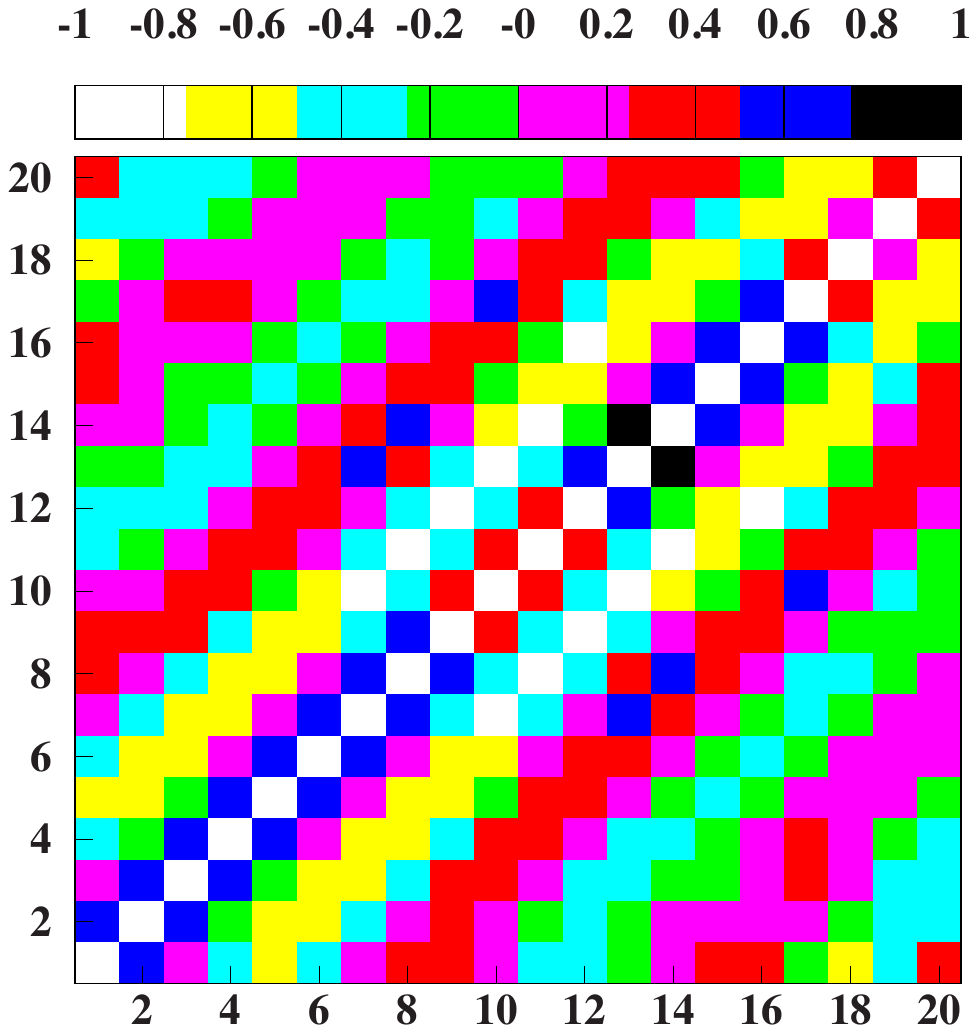}
	\end{subfigure}
   \hfill 
	\begin{subfigure}{0.495\linewidth}
               \includegraphics[width=\linewidth]{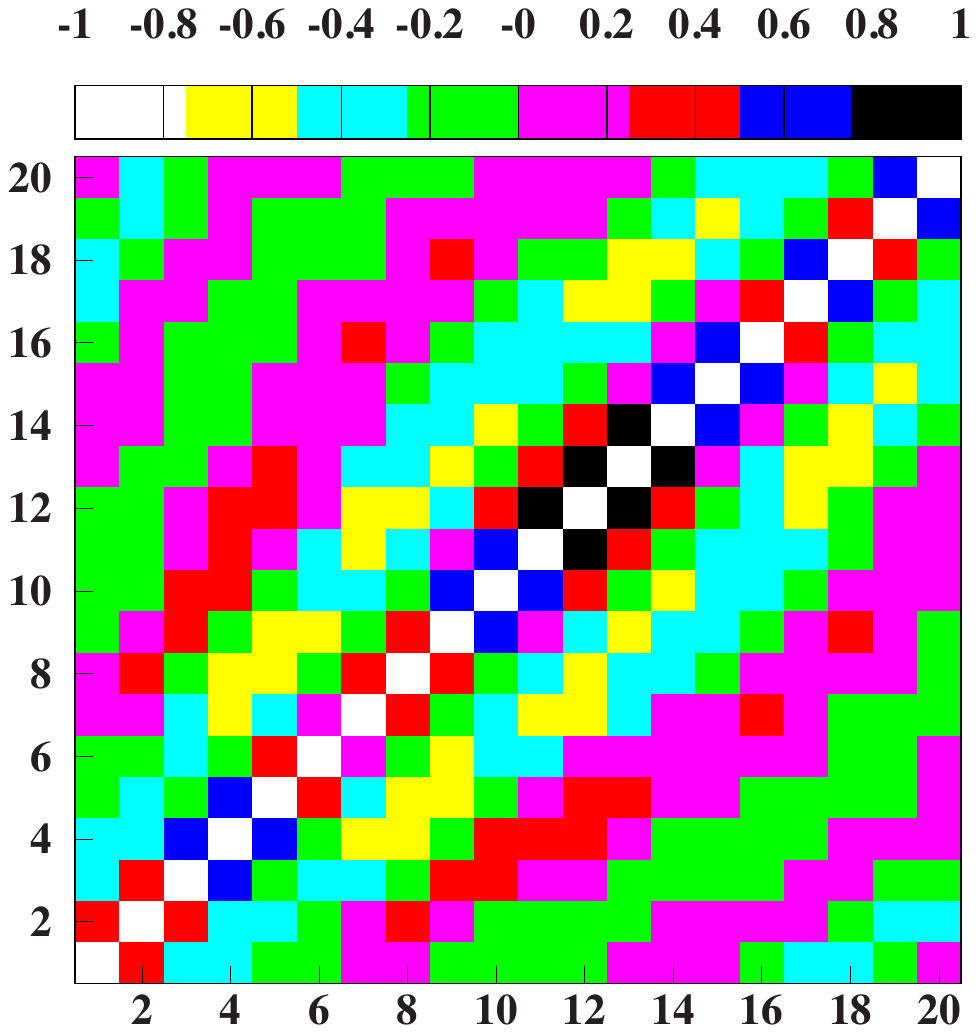}
	\end{subfigure}     
\vspace *{-1cm} 
	\caption{Correlation matrix for RL method  (left) and  correlation matrix for NG method (right).}
	\label{fig:res3}
\end{figure}
\begin{figure}[ht]
	\centering
 	\begin{subfigure}{0.495\linewidth}
		\includegraphics[width=\linewidth]{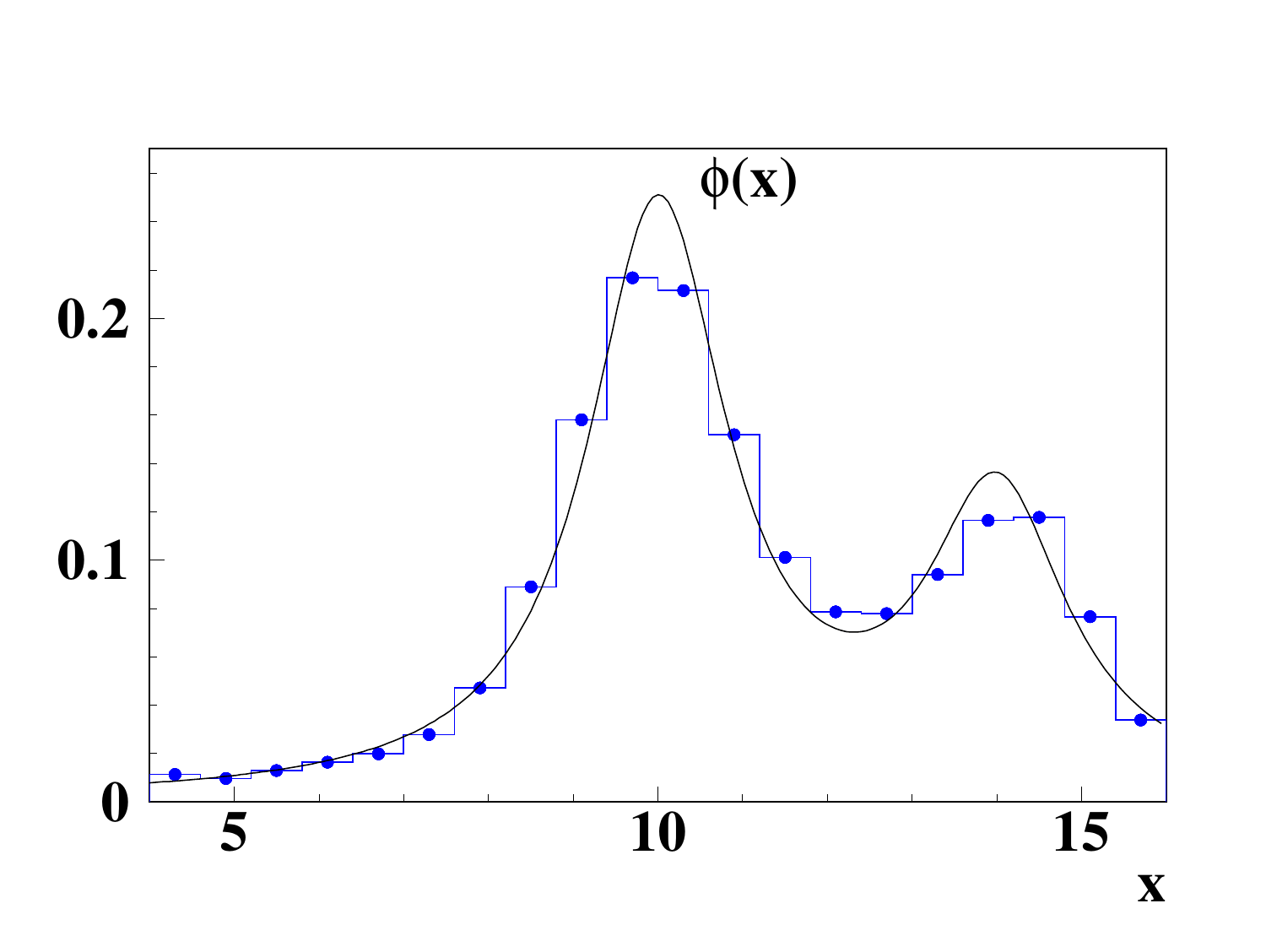}
	\end{subfigure}
   \hfill 
	\begin{subfigure}{0.495\linewidth}
		\includegraphics[width=\linewidth]{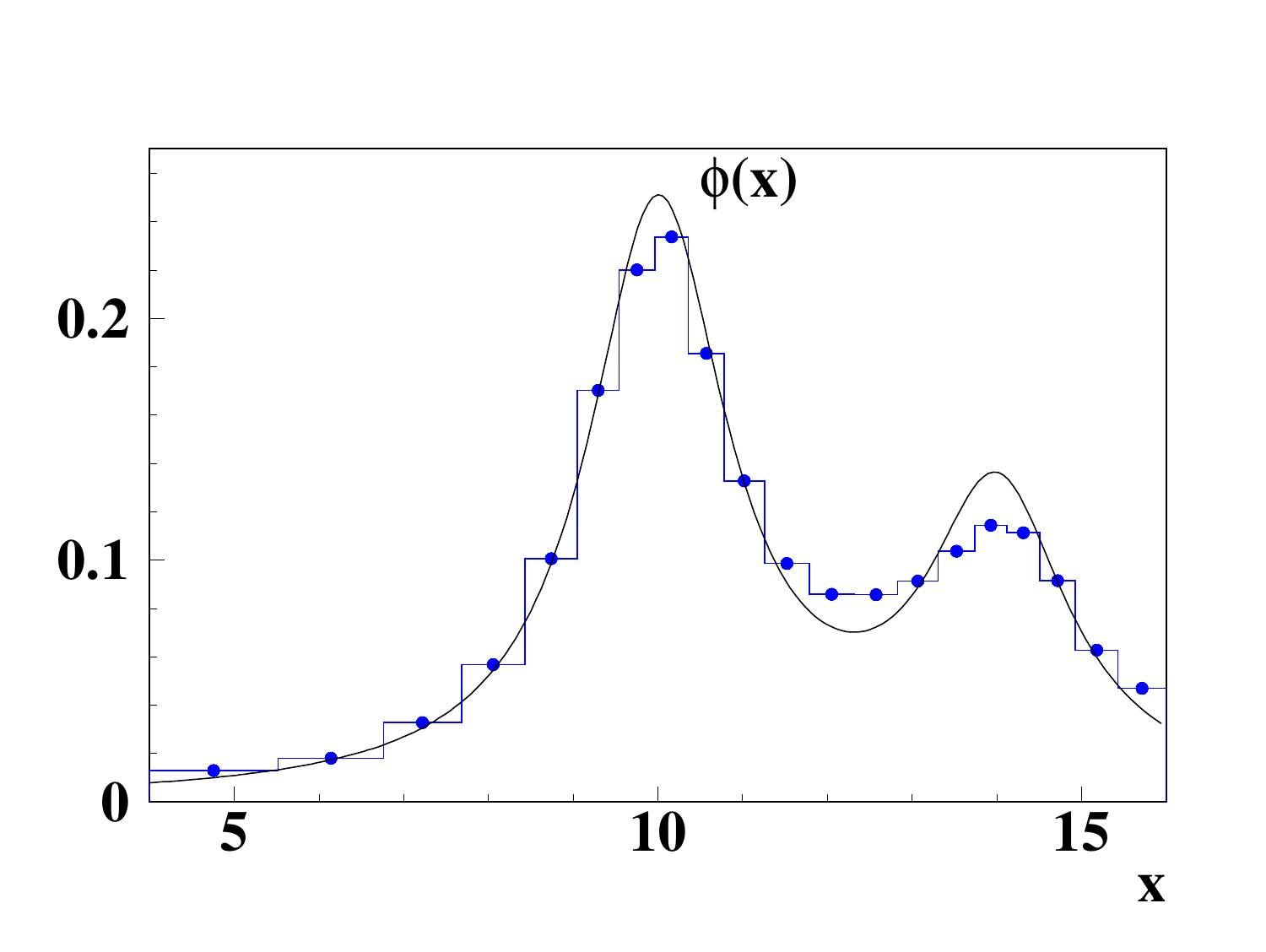}
	\end{subfigure}        
	\caption{Average unfolded PDF for the RL method with the minimum MISE value (left) and  the NG method with the minimum MISE value (right).   For bin $i$, the error bar represents the standard deviation of the estimate of the average value of  $\hat{\phi_i}/(a_{i+1}-a_{i})$. }
	\label{fig:res41}
\end{figure}
\begin{figure}[ht]
	\centering
 	\begin{subfigure}{0.495\linewidth}
		\includegraphics[width=\linewidth]{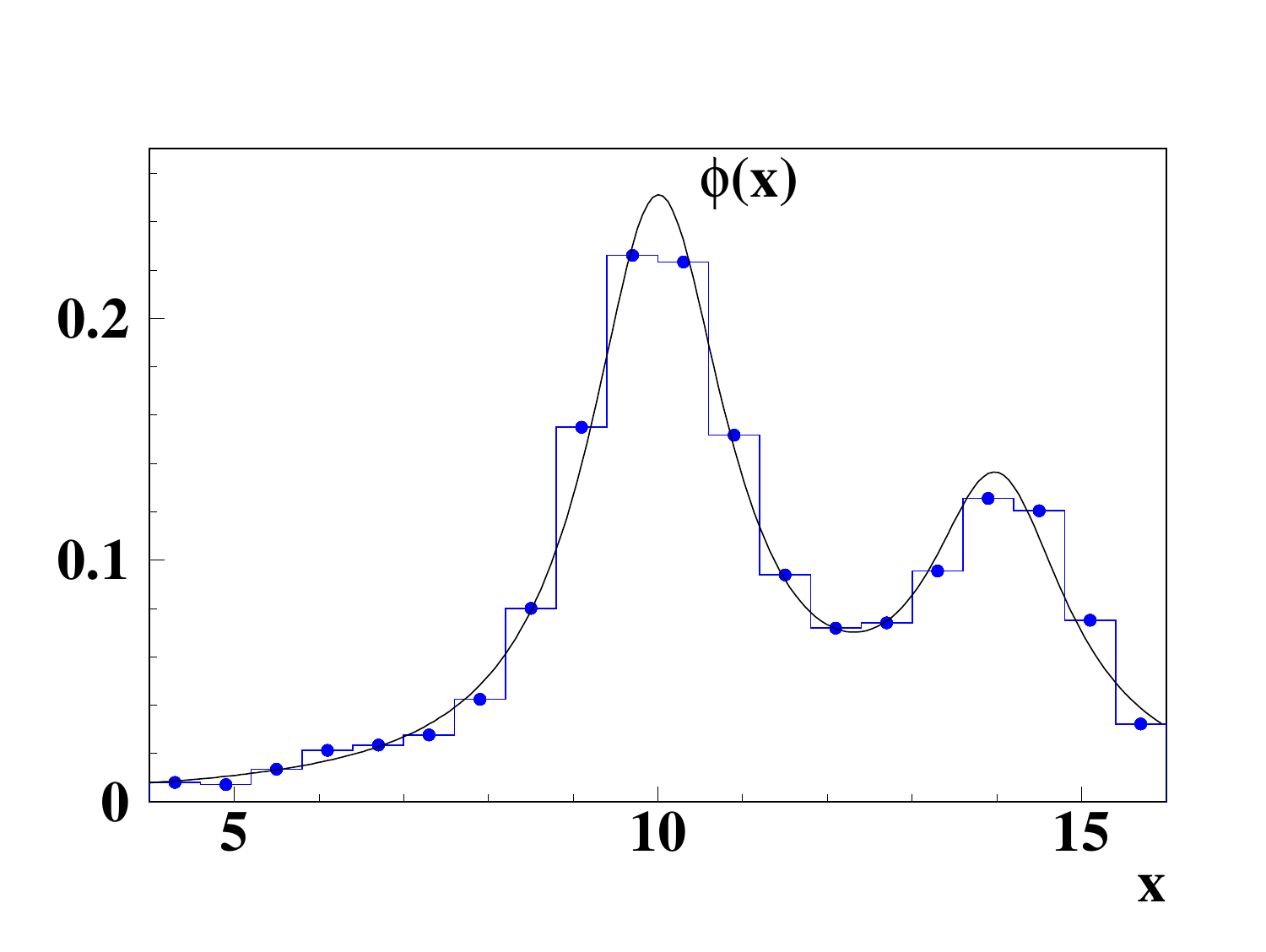}
	\end{subfigure}
   \hfill 
	\begin{subfigure}{0.495\linewidth}
		\includegraphics[width=\linewidth]{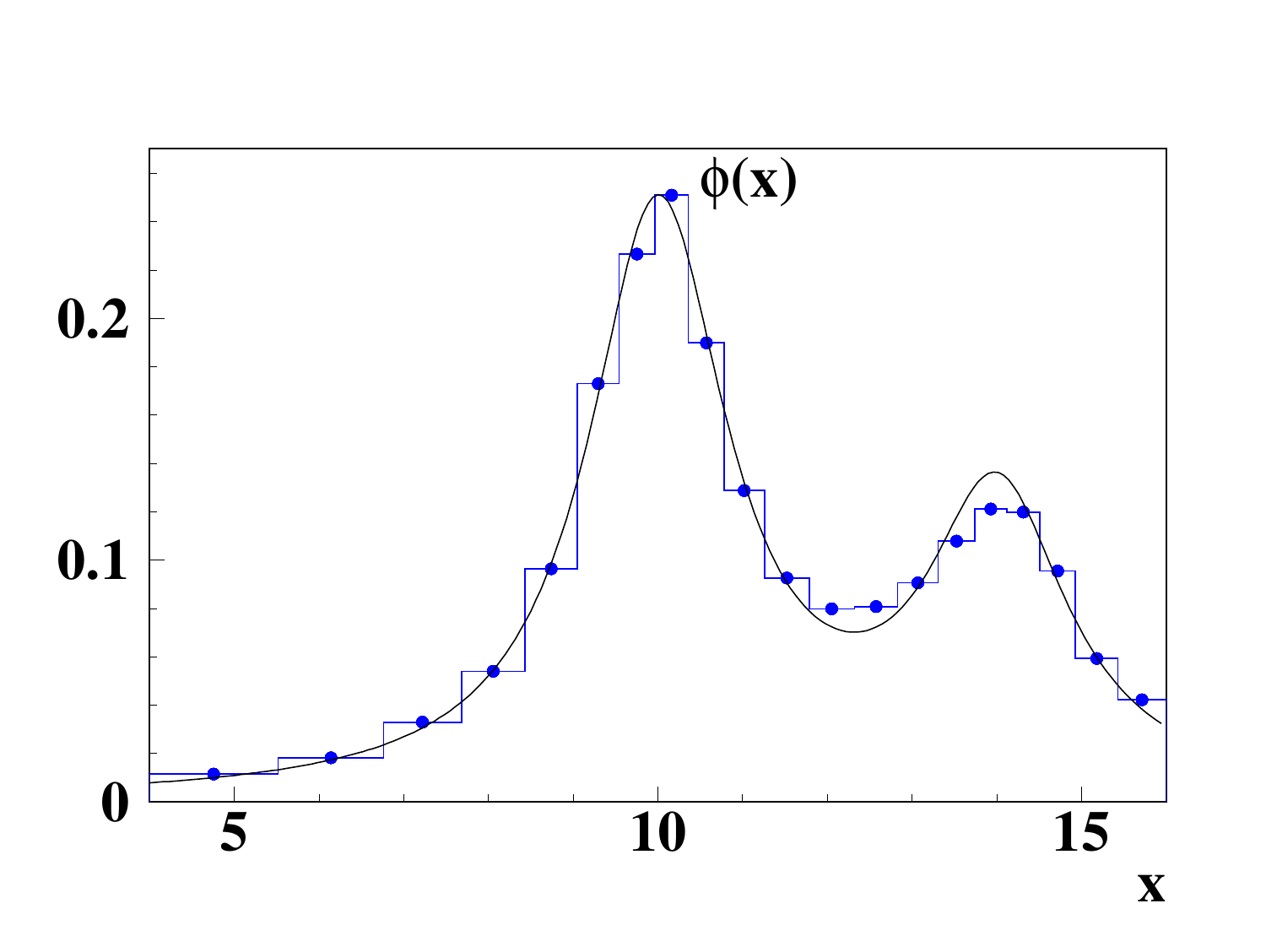}
	\end{subfigure}        
	\caption{Average unfolded PDF for RL method with the minimum MCN value (left) and  average unfolded PDF  for NG method with the minimum MCN value (right).For bin $i$, the error bar represents the standard deviation of the estimate of the average value of  $\hat{\phi_i}/(a_{i+1}-a_{i})$. } 
	\label{fig:res42}
\end{figure}

\newpage
\vspace *{-1cm}
Figure~\ref{fig:res2} presented the unfolded PDF for one sample, calculated using regularization parameters $n=106$ and $\lambda=16.19$, which provide the minimum  value of MISE. Figure~\ref{fig:res3} shows the  correlation matrices for the  RL and NG methods. Figure ~\ref{fig:res41} presents the  average unfolded PDF corresponding to the  optimal value of MISE. Figure ~\ref{fig:res42} shows the  average unfolded PDF for regularization parameters $n=262$ and $\lambda=5.72$,  which provide the minimum value of MCN. 
\vspace *{-0.5cm}

\subsection{Unfolding results for the case of 1000 events}
Fifteen  equidistant bins are defined for the unfolded distribution and 25 equidistant bins for the histogram of reconstructed events in the RL case. Using K-means clustering, 15 bins are defined for the unfolded distribution and 25 bins for the histogram of reconstructed events in the case of the NG method.

\begin{figure}[h]
	\centering
 	\begin{subfigure}{0.495\linewidth}
		\includegraphics[width=\linewidth]{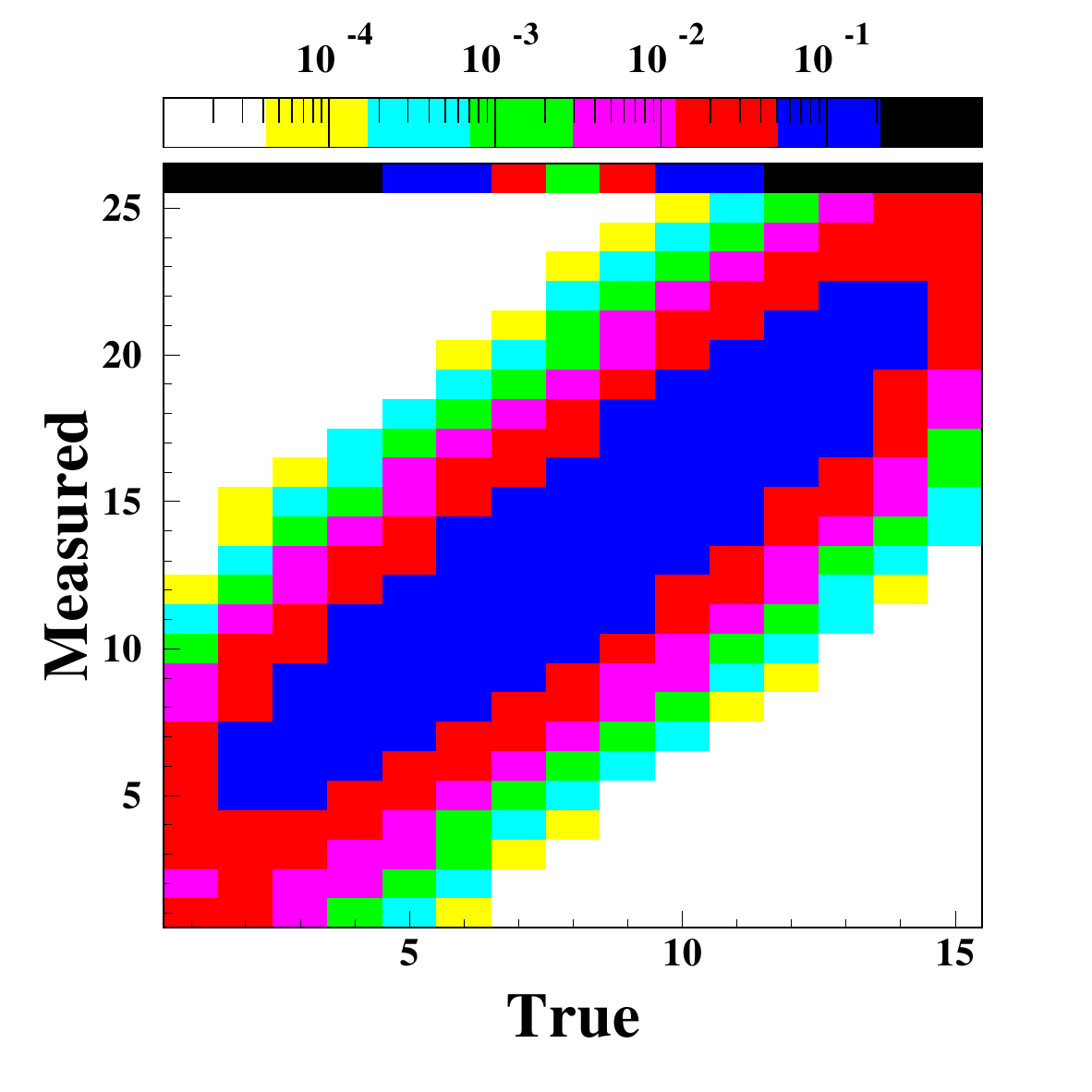}
	\end{subfigure}
   \hfill 
	\begin{subfigure}{0.495\linewidth}
		\includegraphics[width=\linewidth]{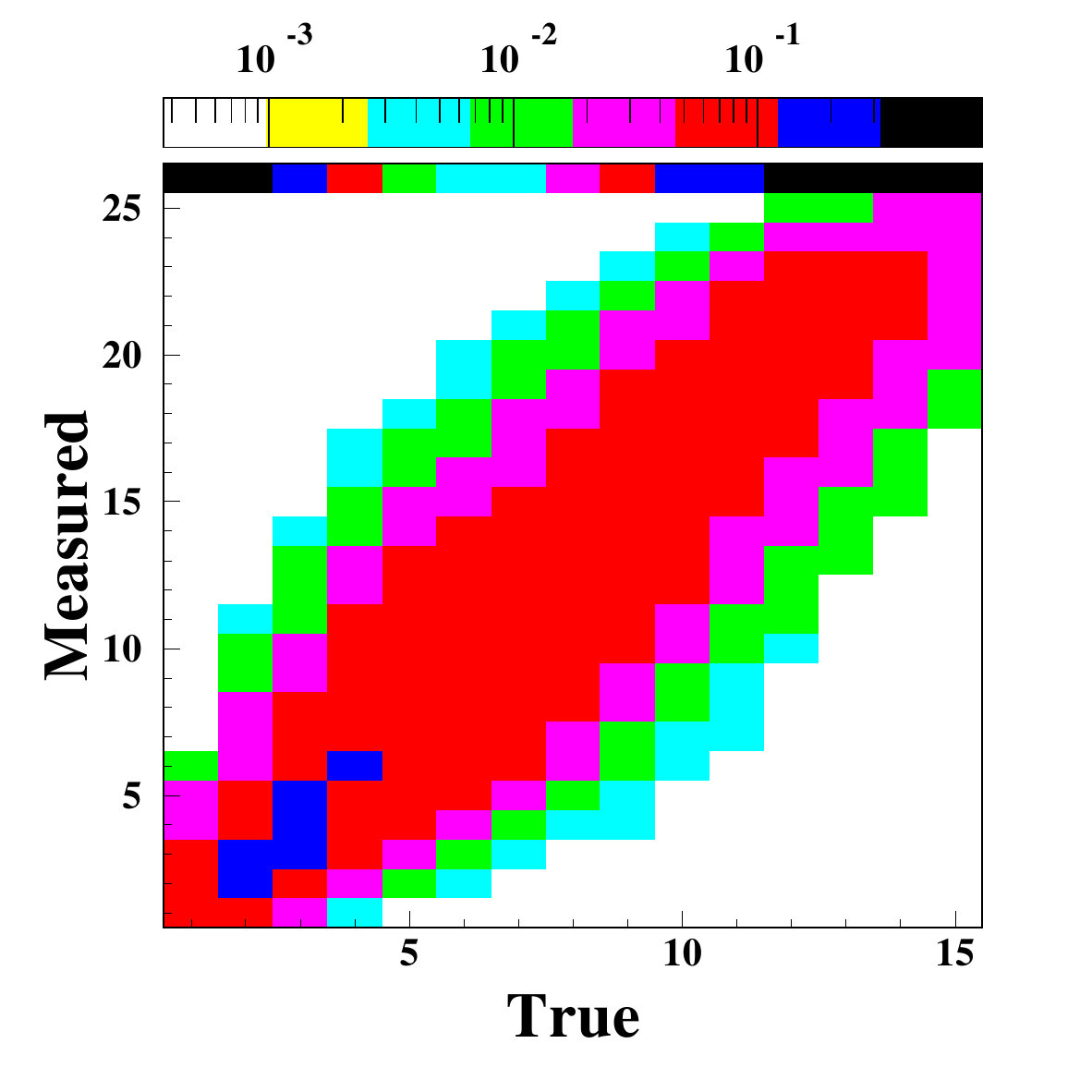}
	\end{subfigure}        
         \caption{ Matrix $R$ calculated for RL method (left) and NG method (right).}  
	\label{fig:rmatr2}
\end{figure}

\begin{figure}[h]
	\centering
 	\begin{subfigure}{0.495\linewidth}
		\includegraphics[width=\linewidth]{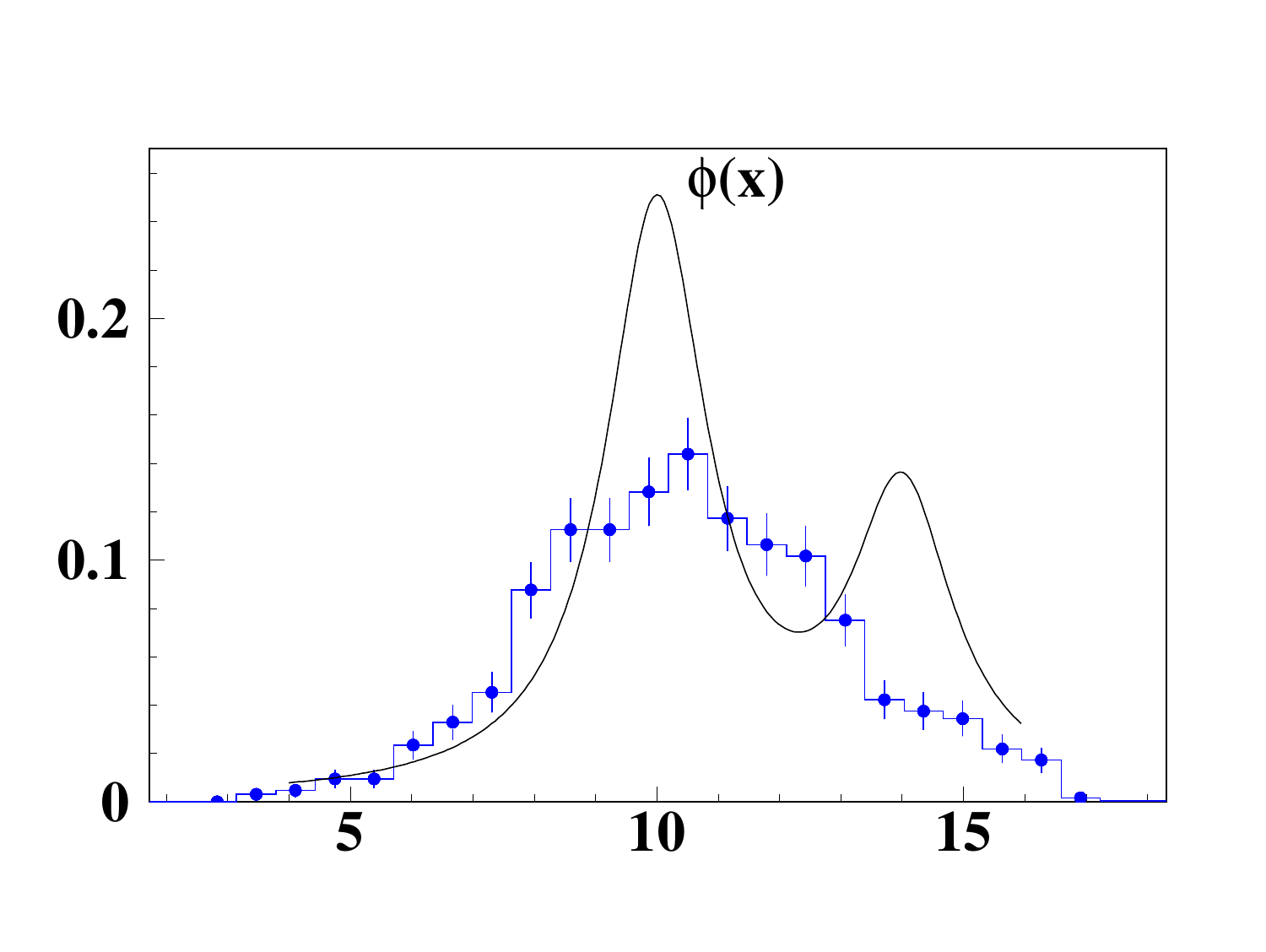}
	\end{subfigure}
   \hfill 
	\begin{subfigure}{0.495\linewidth}
		\includegraphics[width=\linewidth]{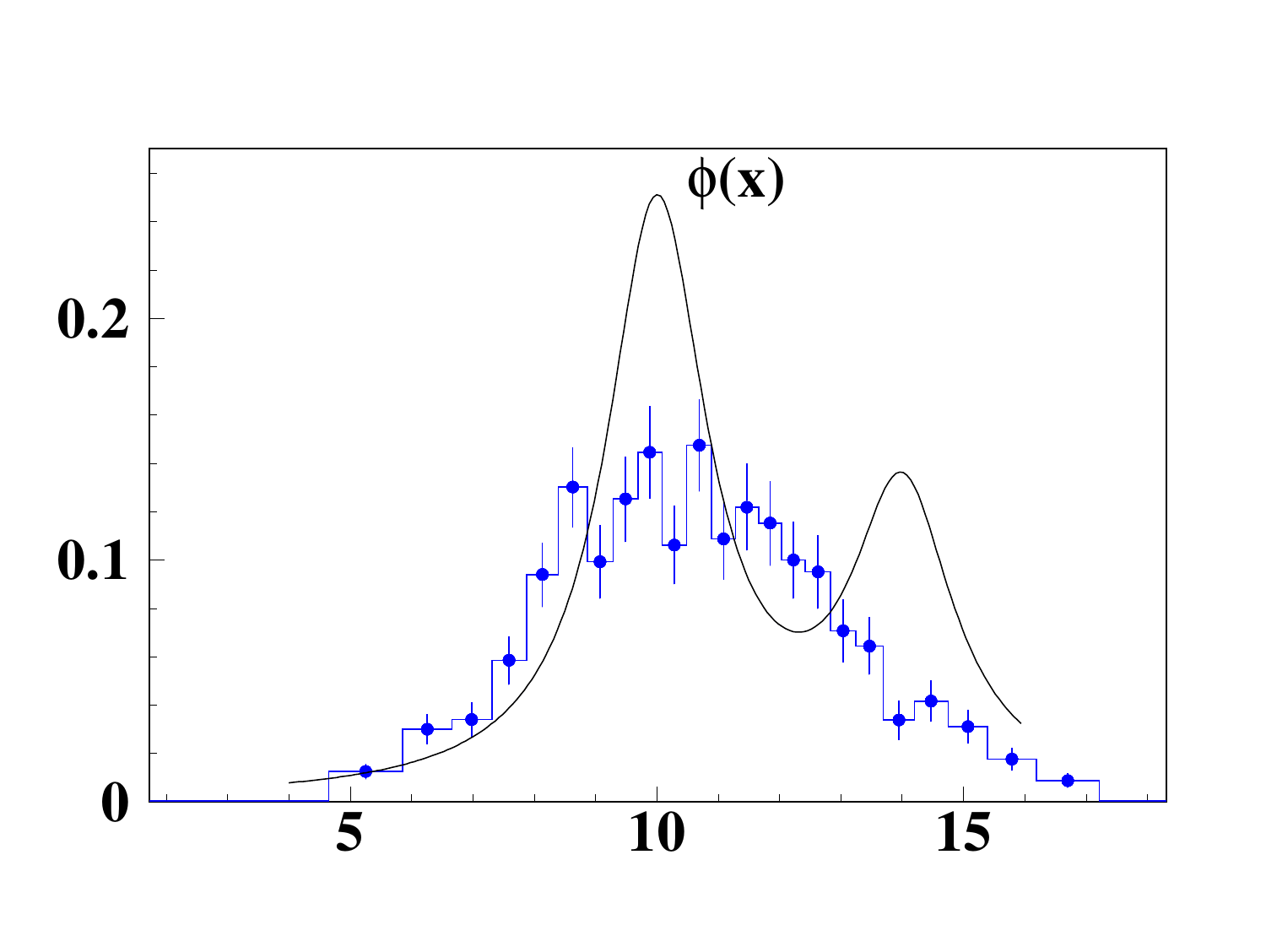}
	\end{subfigure}        
	\caption{Measured distribution histogram for the RL method (left)  and measured distribution histogram for NG method (right). The true distribution  is shown  by the curve.}
	\label{fig:meas2}
\end{figure}

\begin{figure}[h]
	\centering
 	\begin{subfigure}{0.495\linewidth}
		\includegraphics[width=\linewidth]{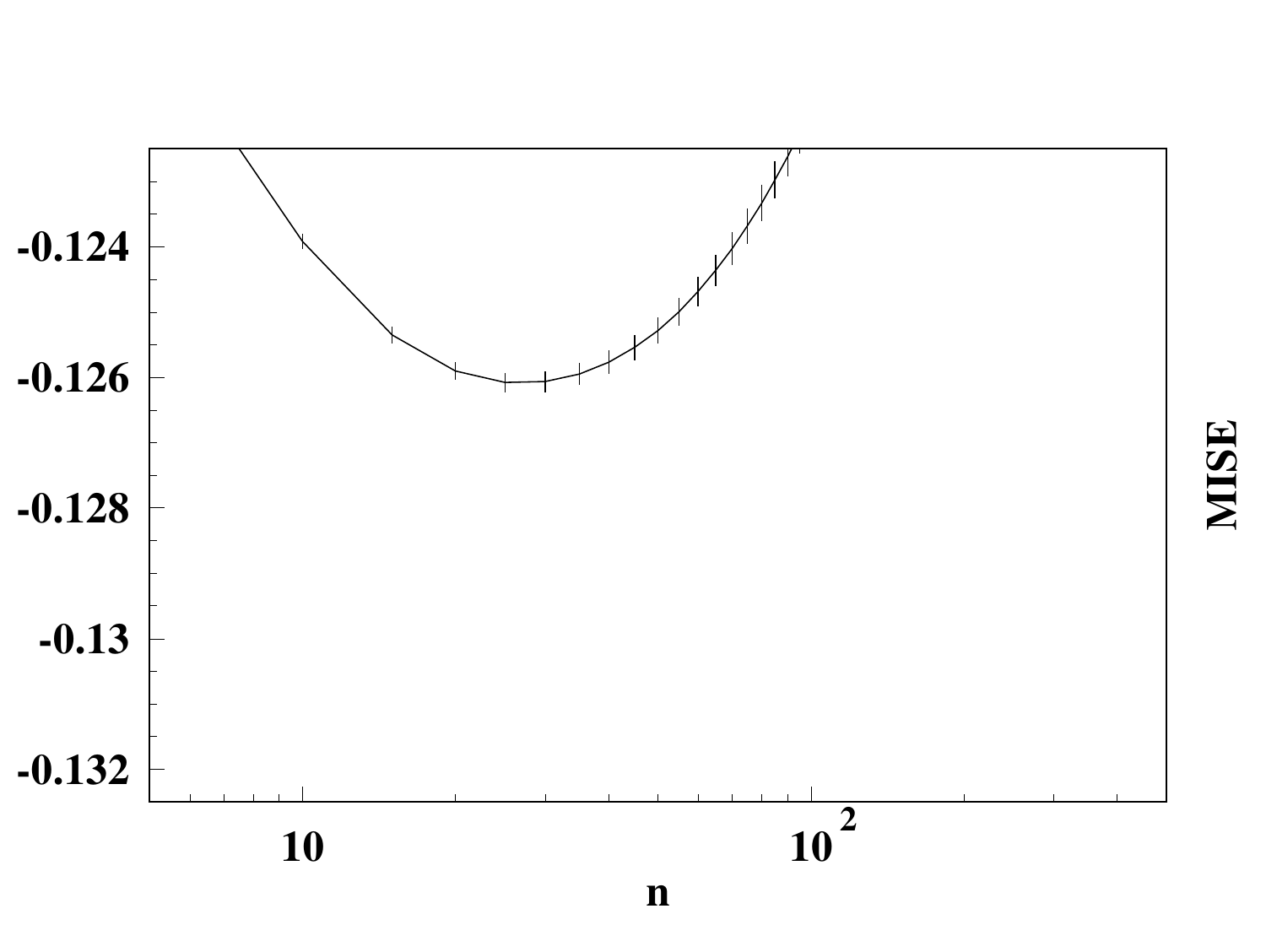}
	\end{subfigure}
   \hfill 
	\begin{subfigure}{0.495\linewidth}
		\includegraphics[width=\linewidth]{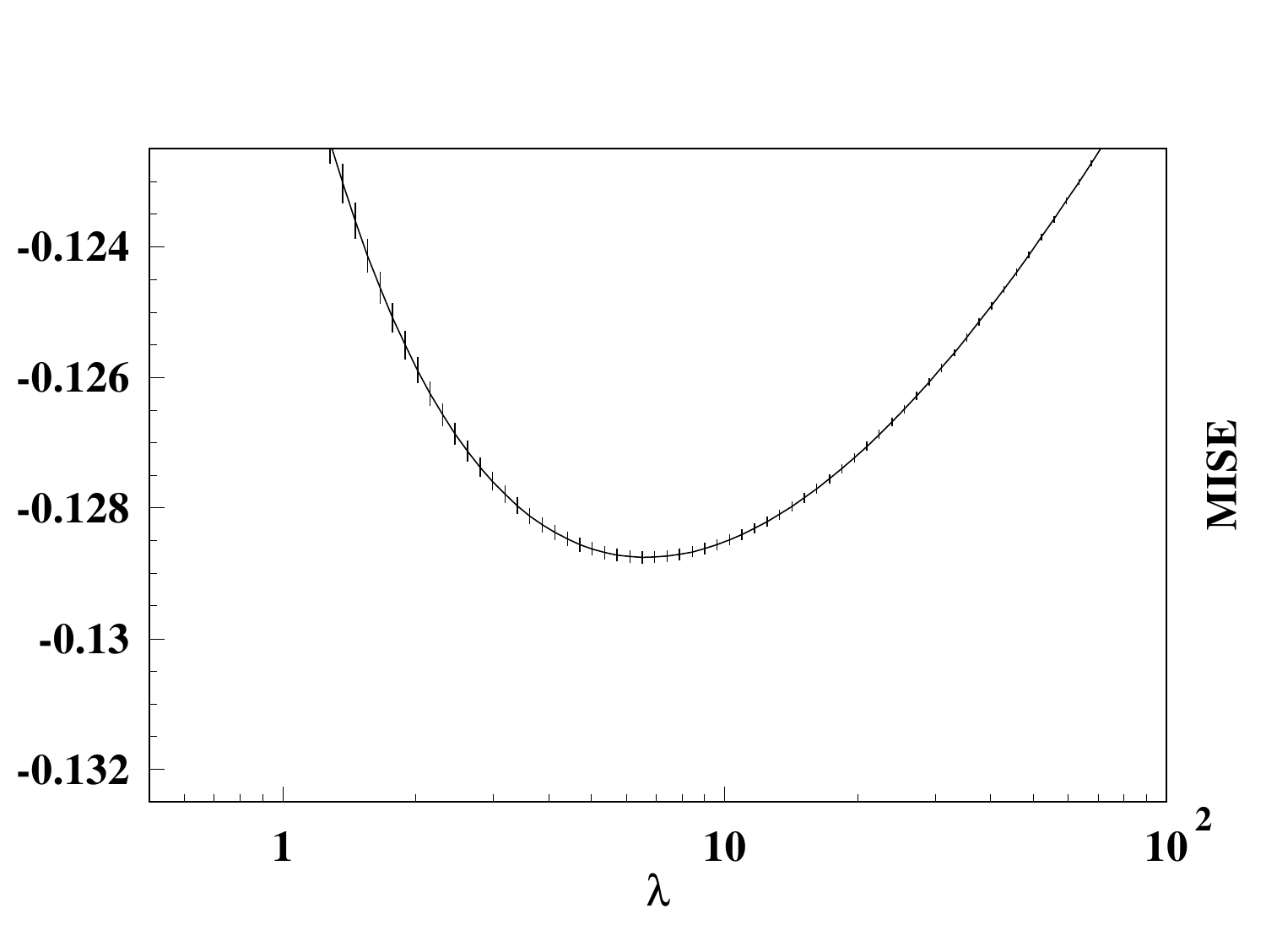}
	\end{subfigure}        
	\caption{MISE for RL method (left) and MISE for the NG method(right).  The standard deviations of the estimated MISE are presented as error bars.
  }
	\label{fig:mise2}
\end{figure}

\begin{figure}[h]
	\centering
 	\begin{subfigure}{0.495\linewidth}
		\includegraphics[width=\linewidth]{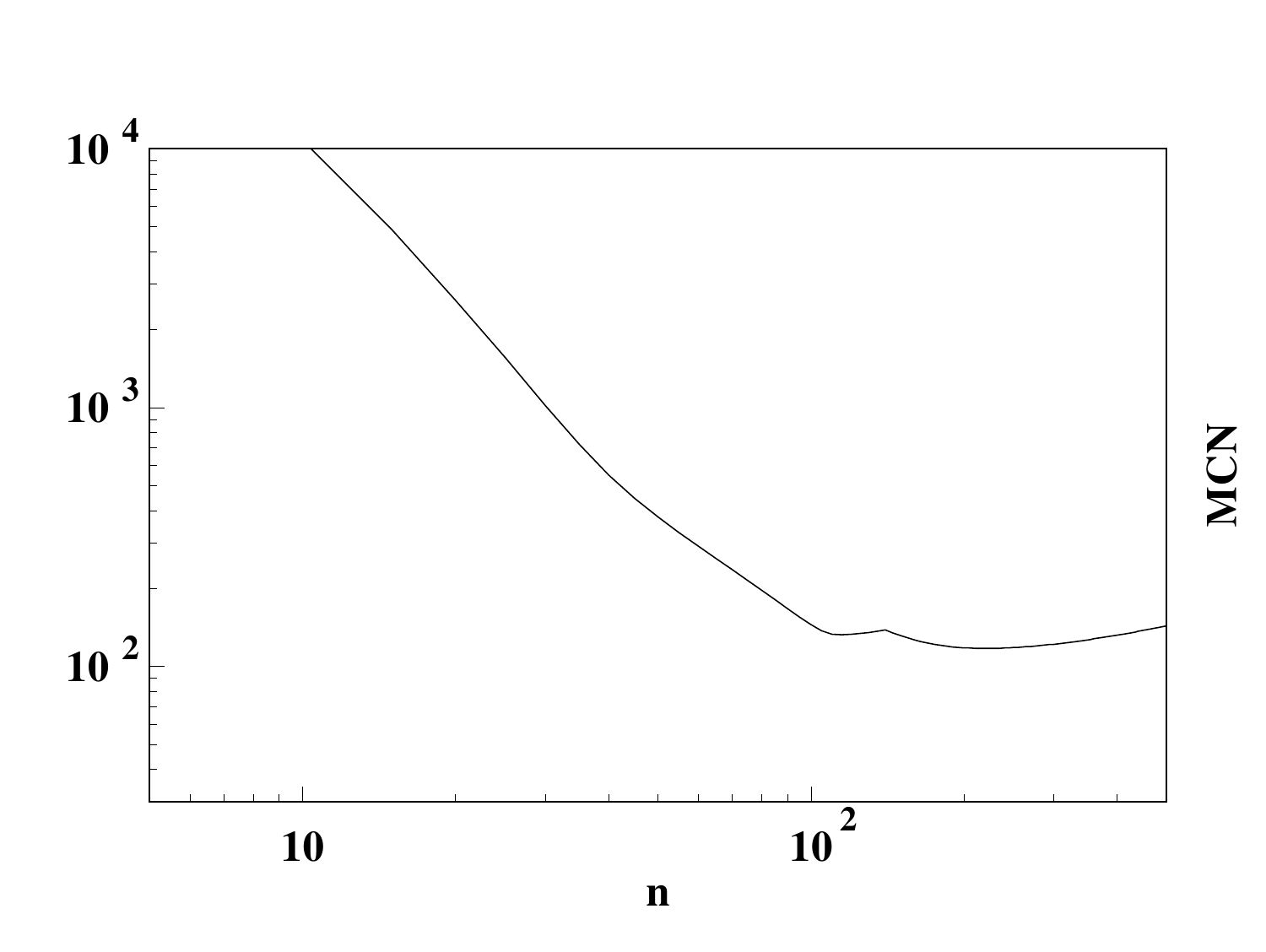}
	\end{subfigure}
   \hfill 
	\begin{subfigure}{0.495\linewidth}
		\includegraphics[width=\linewidth]{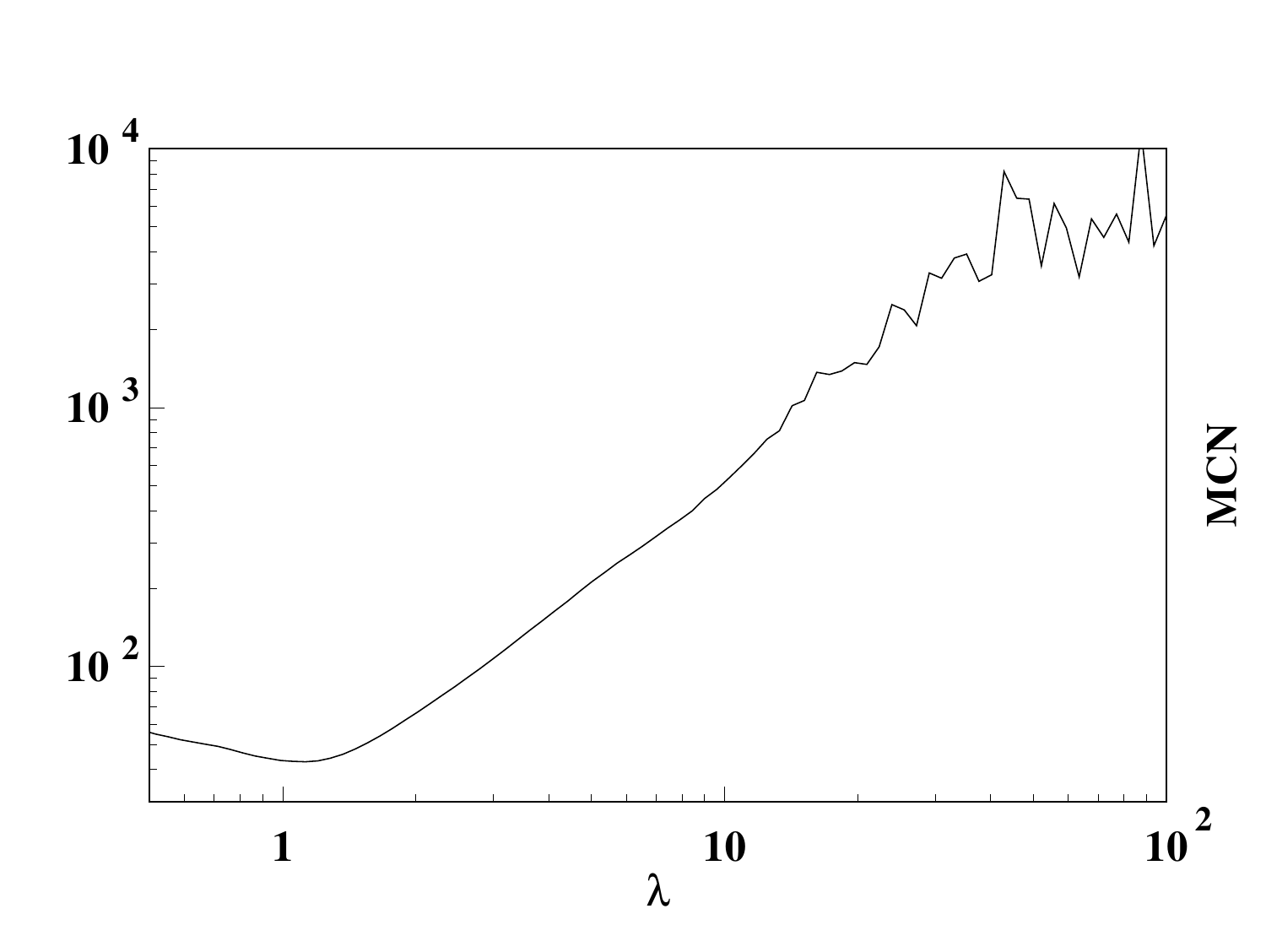}
	\end{subfigure}        
	\caption{MCN for RL method (left) and  MCN for NG method (right).  }
	\label{fig:cond2}
\end{figure}

The response matrices are presented in Figure~\ref{fig:rmatr2} , and the histograms of measured distributions are shown in Figure~\ref{fig:meas2}. To determine the minimum values of the MISE and of the MCN, the regularization parameter was scanned in both cases. The results of the calculations are presented in Figure~\ref{fig:mise2} and  \ref{fig:cond2}. Five hundred replications of the unfolded distribution were used for this analysis.

\newpage
\begin{figure}[ht]
	\centering
 	\begin{subfigure}{0.495\linewidth}
		\includegraphics[width=\linewidth]{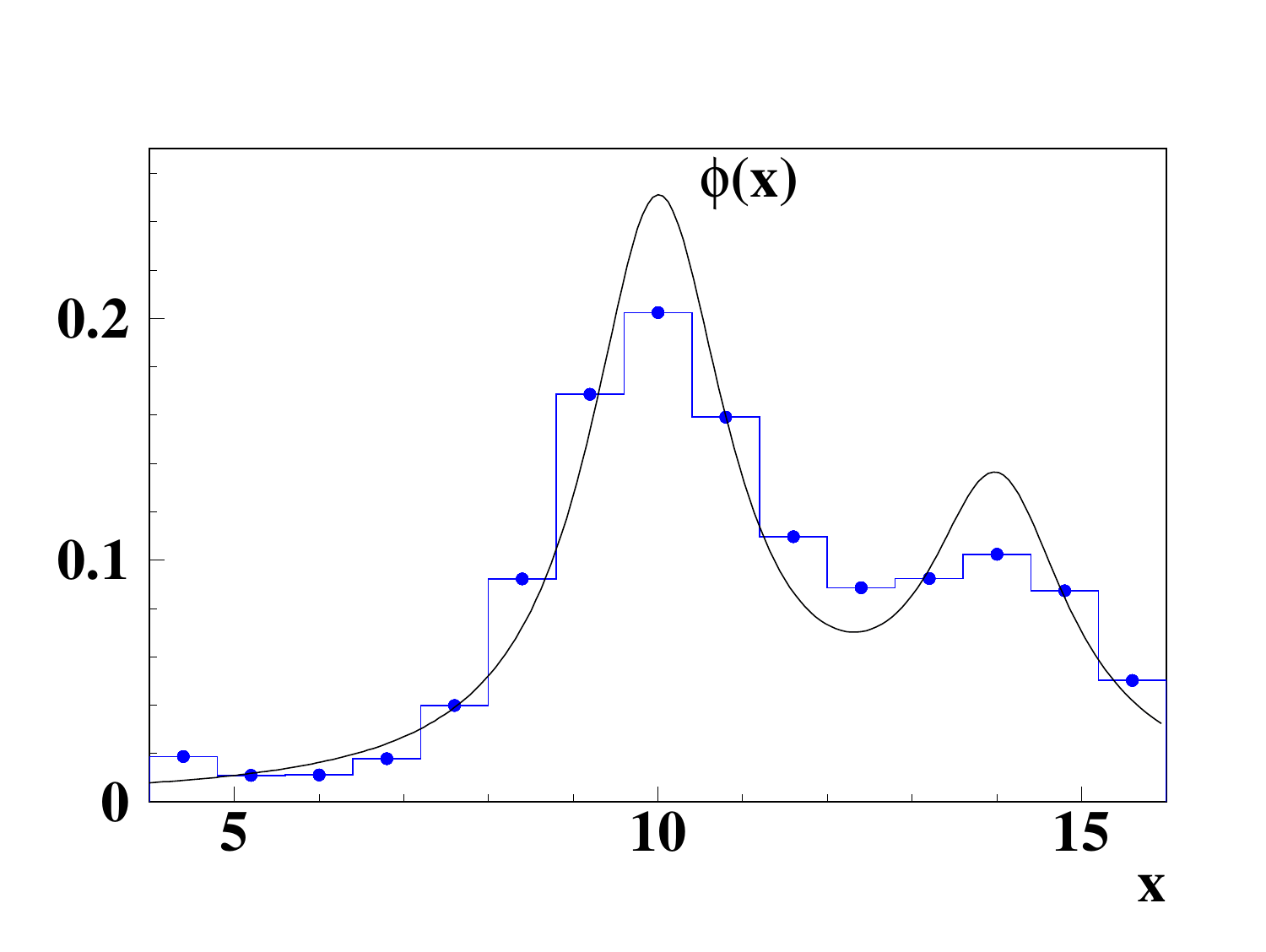}
	\end{subfigure}
   \hfill 
	\begin{subfigure}{0.495\linewidth}
		\includegraphics[width=\linewidth]{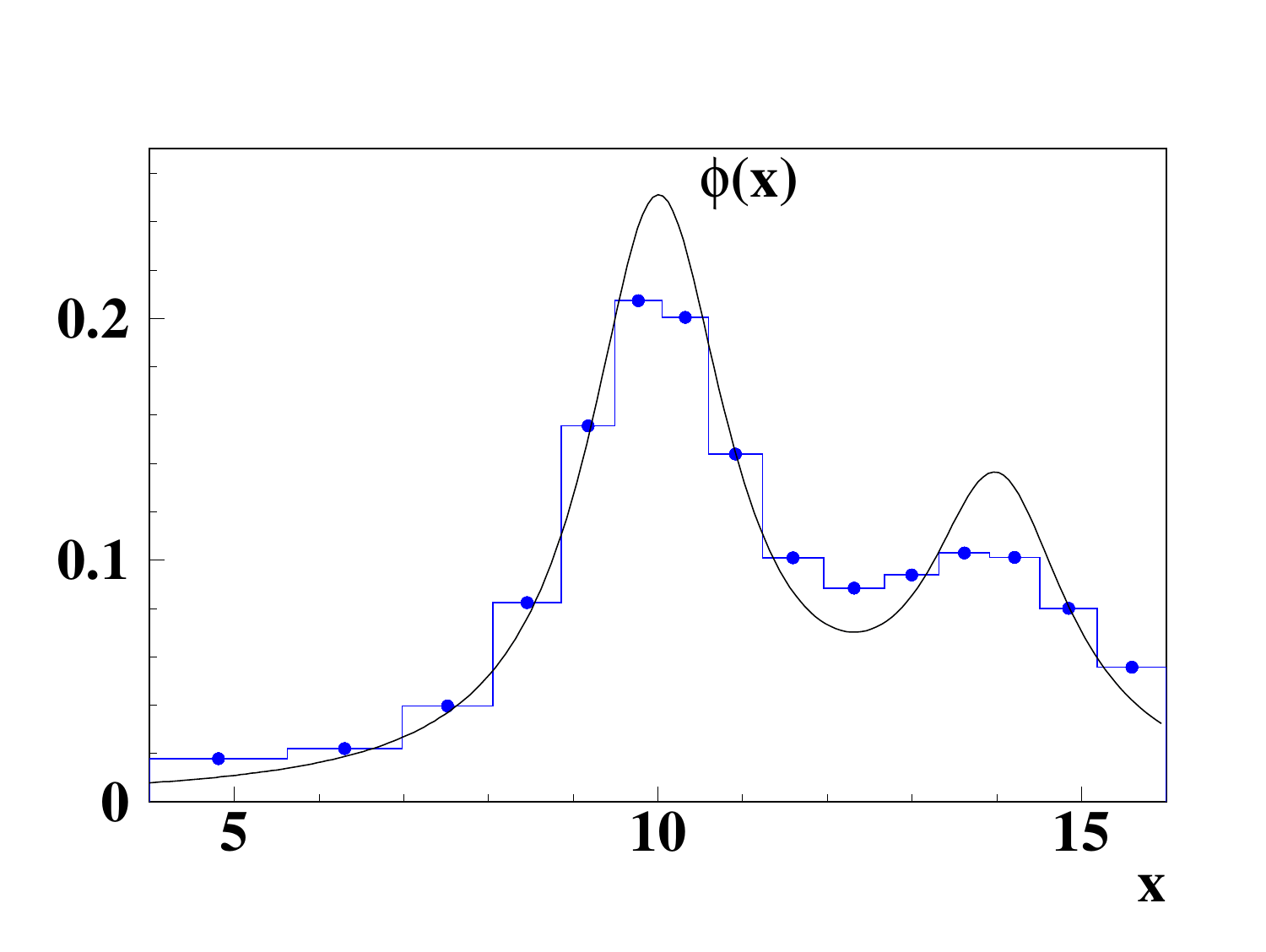}
	\end{subfigure}        
	\caption{Average unfolded PDF for the RL method with the minimum MISE value (left) and  the NG method with the minimum MISE value (right).   For bin $i$, the error bar represents the standard deviation of the estimate of the average value of  $\hat{\phi_i}/(a_{i+1}-a_{i})$. }
	\label{fig:res4aa}
\end{figure}

\begin{figure}[ht]
	\centering
 	\begin{subfigure}{0.495\linewidth}
		\includegraphics[width=\linewidth]{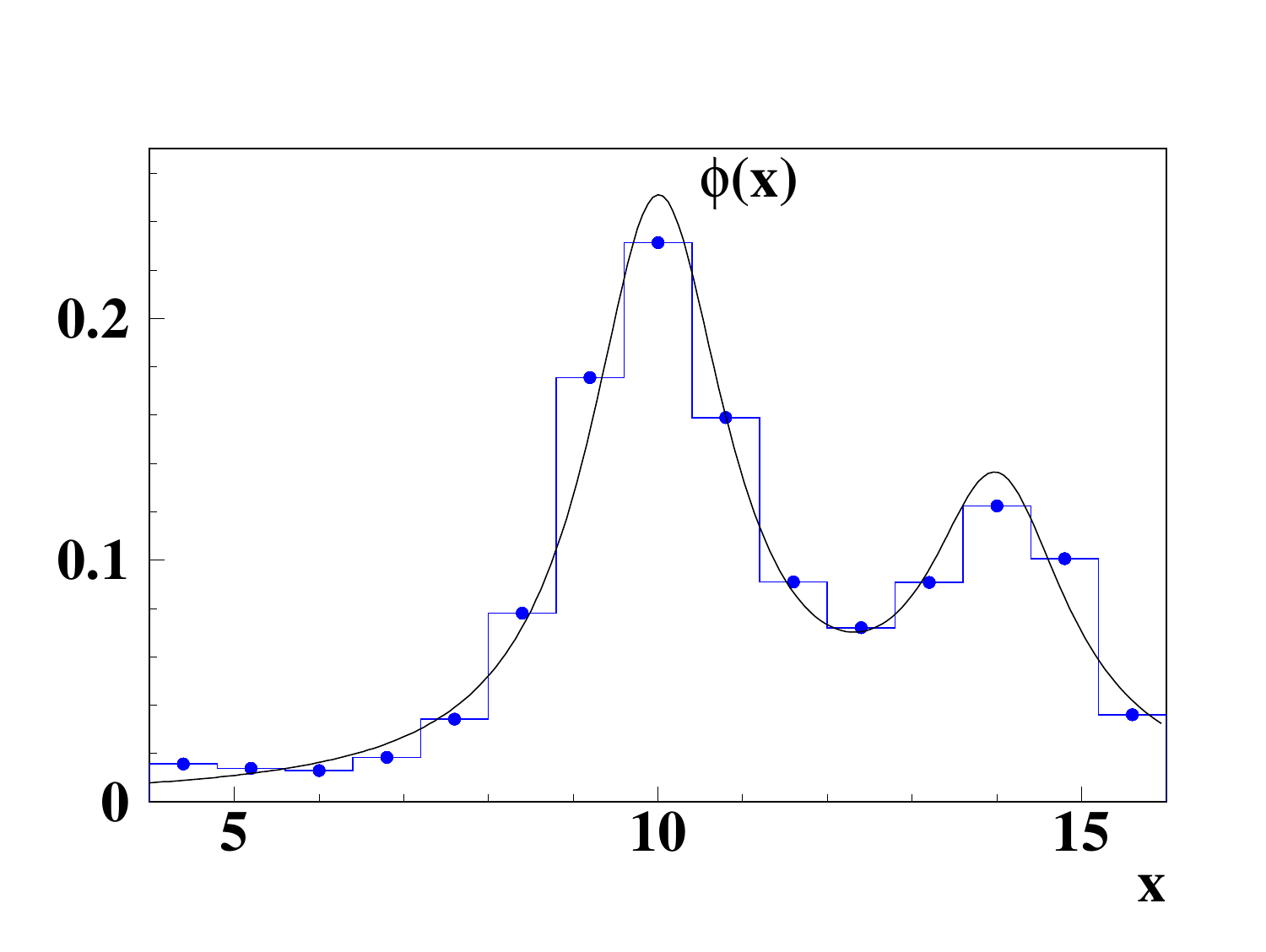}
	\end{subfigure}
   \hfill 
	\begin{subfigure}{0.495\linewidth}
		\includegraphics[width=\linewidth]{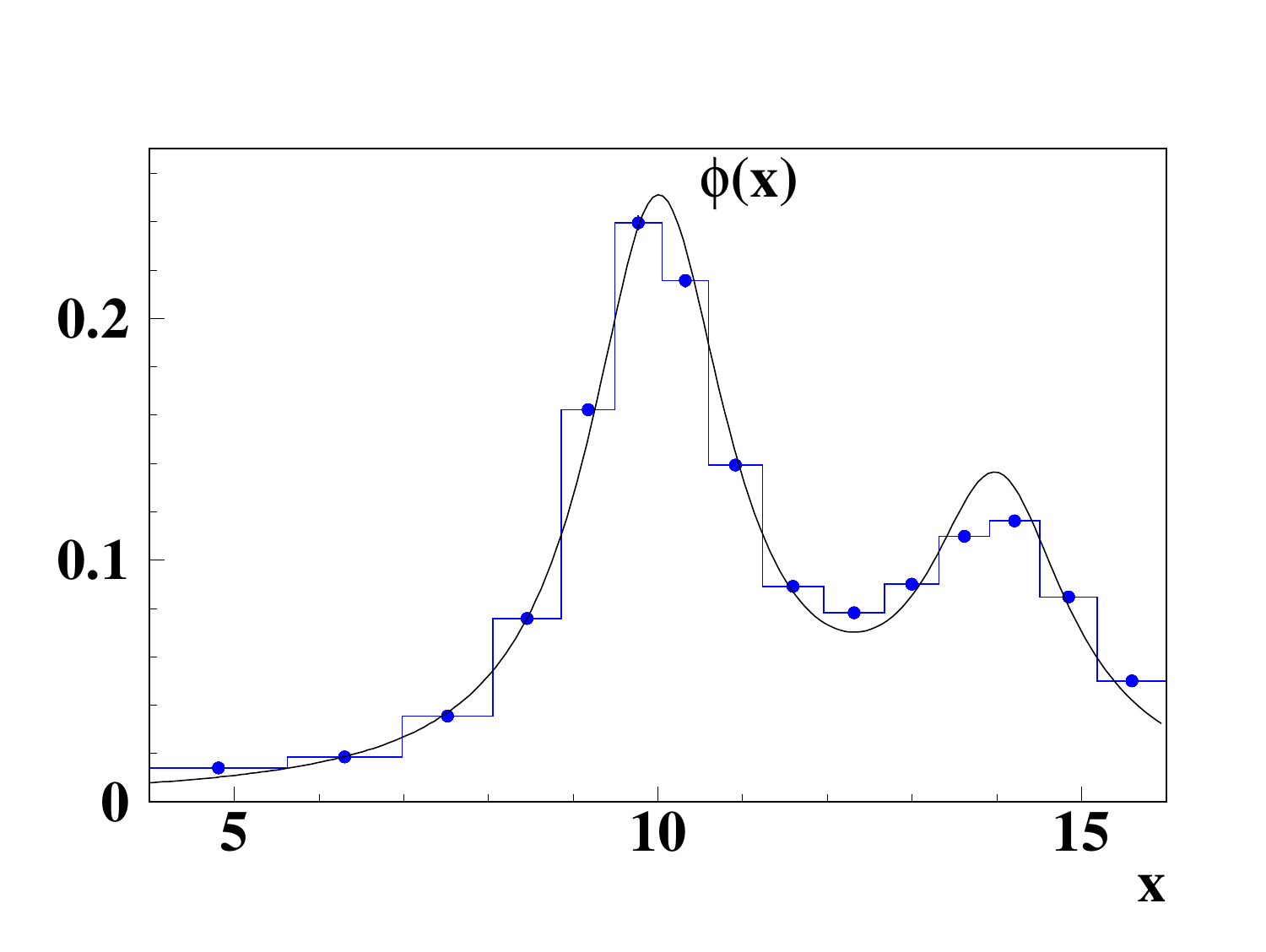}
	\end{subfigure}        
      \caption{Average unfolded PDF for RL method with the minimum MCN value (left) and  average unfolded PDF  for NG method with the minimum MCN value (right).  For bin $i$, the error bar represents the standard deviation of the estimate  of the average value of  $\hat{\phi_i}/(a_{i+1}-a_{i})$. } 

	\label{fig:res44aa}
\end{figure}

\begin{center}
\captionof{table}{Characteristics of the minima of MISE and  MCN for the  RL  and NG algorithms for case of 1000 measured events.}
\begin{tabular}{ | c | l l  l | l  l  l |  }
\hline
& \multicolumn{3}{ c }{Minimum MISE }  & \multicolumn{3} {c|} {Minimum MCN} \\
\hline
 Method &MISE & MCN & Reg. par. &MISE & MCN & Reg. par. \\
\hline 
  RL & -0.12606& 1302.6 &27  & -0.11354&117.3&221\\
  NG & -0.12876&291.1 & 6.52 &-0.12101 & 42.8 & 1.13\\
\hline
\end{tabular}
\end{center}
According  to the data presented in Table 2, the NG method exhibits a lower value of the optimal MISE in both cases. It also has a lower MCN compared to the RL method, indicating that the NG method handles multicollinearity more effectively. This suggests that the NG method is likely to be more robust and produce more stable results.
\vspace{-2cm}
\begin{figure}
	\centering
 	\begin{subfigure}{0.495\linewidth}
		\includegraphics[width=\linewidth]{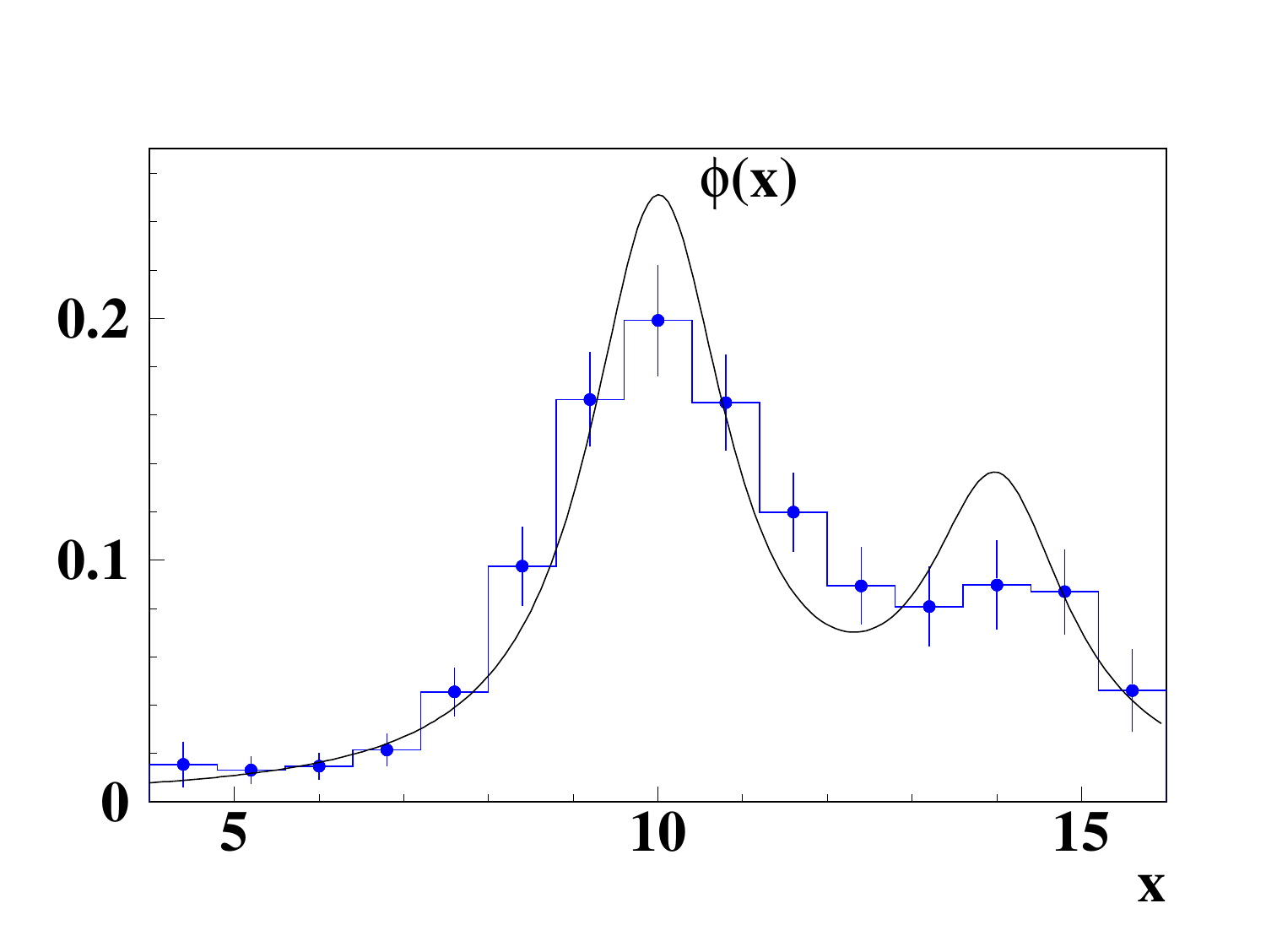}
	\end{subfigure}
   \hfill 
	\begin{subfigure}{0.495\linewidth}
		\includegraphics[width=\linewidth]{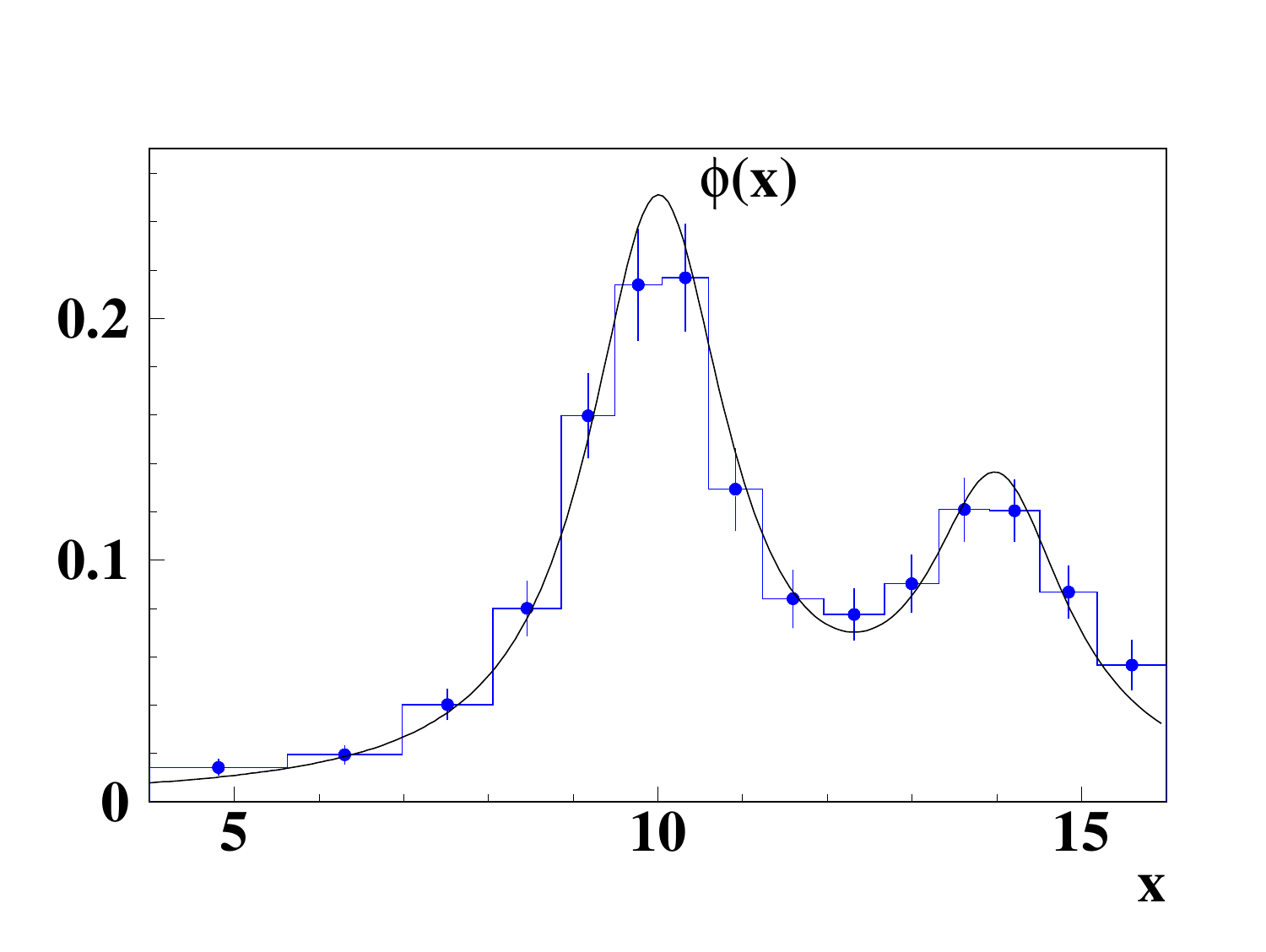}
	\end{subfigure}        
	\caption{Unfolded PDF  of the RL method with the minimum MISE value (left)  and the unfolded PDF of the NG method with  the minimum MISE value (right). The true distribution  is represented  by the curve. For bin $i$, the error bar represents the standard deviation  of  $\hat{\phi_i}/(a_{i+1}-a_{i})$.}
	\label{fig:res2dd}
\end{figure}
\begin{figure}
	\centering
 	\begin{subfigure}{0.495\linewidth}
                 \includegraphics[width=\linewidth]{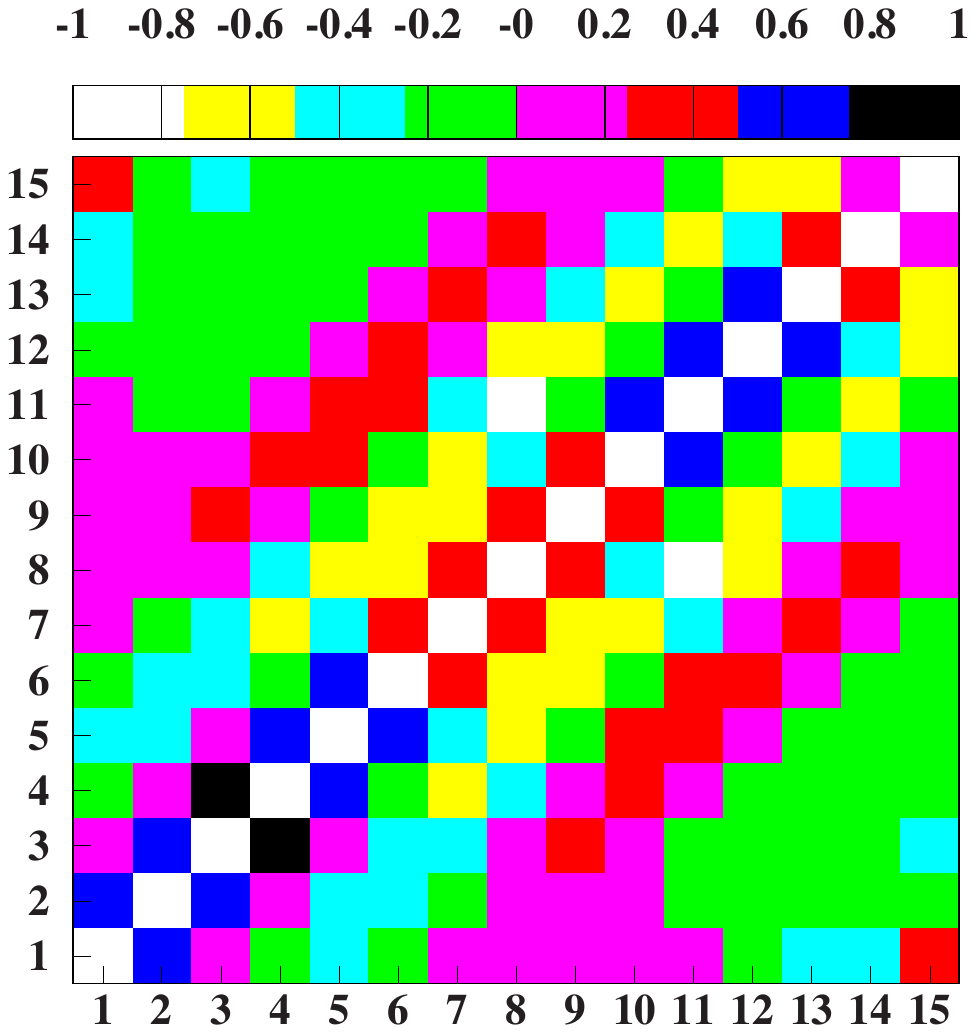}
	\end{subfigure}
   \hfill 
	\begin{subfigure}{0.495\linewidth}
                \includegraphics[width=\linewidth]{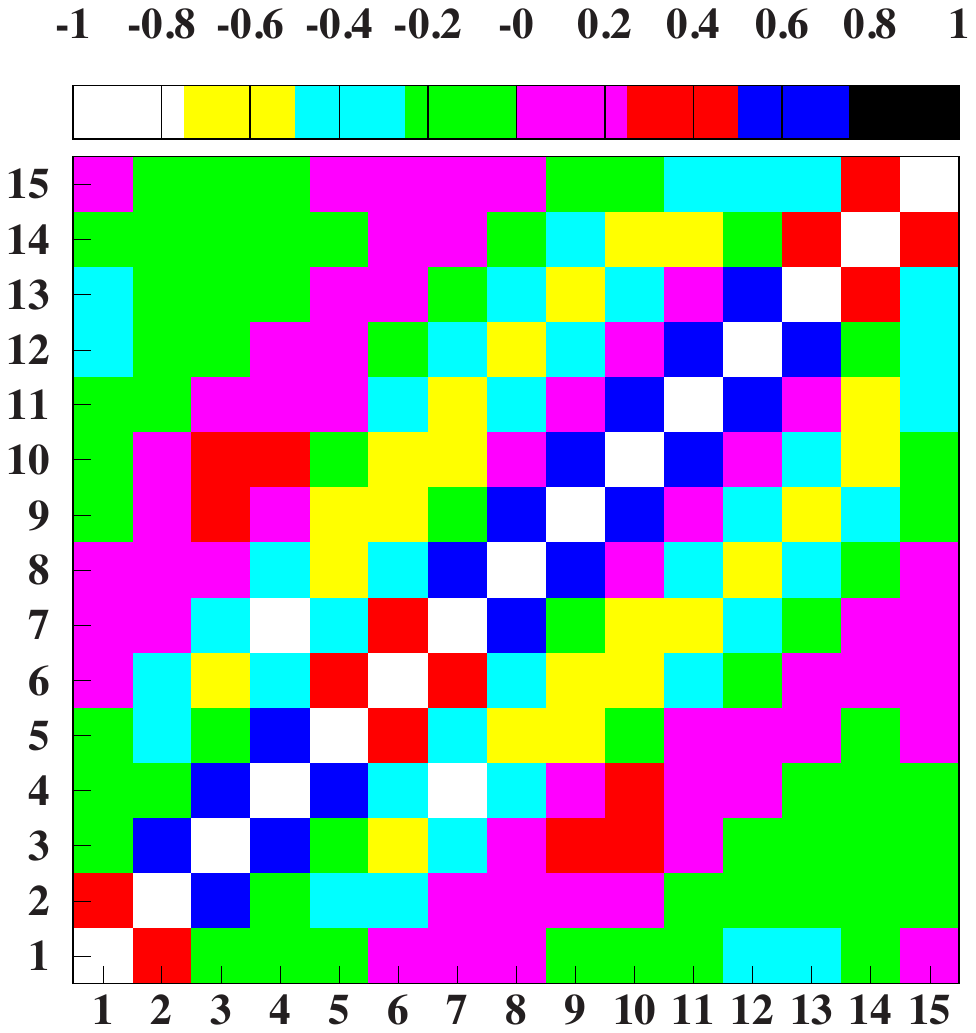}
	\end{subfigure}   
\vspace*{-1cm}     
	\caption{Correlation matrix for RL method  (left) and  correlation matrix for NG method (right).}
	\label{fig:res33}
\end{figure}
\newpage
\vspace *{-1cm}
Figure~\ref{fig:res2dd} presents the unfolded PDF for one sample, calculated using the regularization parameters $n=27$ and $\lambda=6.52$, which provide the minimum value of MISE. Figure~\ref{fig:res33} shows the correlation matrices for the RL and NG methods. Figure ~\ref{fig:res4aa} presents the  average unfolded PDF corresponding to the optimal value of MISE. Figure ~\ref{fig:res44aa} shows the average unfolded PDF for regularization parameters $n=221$ and $\lambda=1.13$,  which provide the minimum  value of MCN.
\section{Conclusions}
In this paper, unfolding is considered as a procedure for estimating the probability density function. According to this definition, the quality of the estimation is defined by the mean integrated square error  and the minimum condition number of the correlation matrix for the set of points representing the estimated distribution. A comparison is made between the Richardson-Lucy method and a new data unfolding method with mean integrated  square error  optimization, which was recently developed by the author. A numerical example demonstrates the clear advantages of the new method. Additionally, it is shown that the newly introduced internal criterion for assessing goodness is a powerful tool for data unfolding.
\section {Acknowledgments}
 The author would like to express his gratitude to the anonymous reviewer for their valuable feedback and constructive comments, which have significantly improved the quality of this manuscript.
\section *{References}
\bibliography{project}
\end{document}